\documentclass[12pt, a4paper]{article}
\pdfoutput=1

\usepackage[utf8]{inputenc} 
\usepackage[T1]{fontenc}

\usepackage{amsmath}
\usepackage{amssymb}
\usepackage{amsfonts}
\usepackage{bbm}
\usepackage{verbatim}
\usepackage{dsfont}
\usepackage{booktabs}
\usepackage{slashed}
\usepackage{bm}
\usepackage{subcaption}
\usepackage{cancel}

\usepackage{mathtools}

\usepackage{epsfig}
\usepackage{color}
\usepackage[table]{xcolor}
\usepackage{graphicx}
\usepackage[]{caption}
\usepackage{float}
\usepackage{overpic}

\usepackage{enumerate}
\usepackage{hhline}
\usepackage{multirow}
\usepackage{xspace}
\usepackage{setspace}
\usepackage[title]{appendix}

\usepackage{soul}
\renewcommand{\theequation}{\arabic{section}.\arabic{equation}}

\newcommand{\F}{{\cal F}}

\newcommand{\A}{{\cal A}}

\newcommand{\V}{{\cal V}}

\newcommand{\cH}{{\cal H}}

\newcommand{\Z}{\mathbb{Z}}

\renewcommand{\Re}{{\rm Re}\,}
\renewcommand{\Im}{{\rm Im}\,}

\renewcommand{\and}{\mbox{and}}

\usepackage{xcolor}

\newcommand{\Nf}{N_{\rm flux}}

\newcommand{\rank}{{\rm rank}\,}

\newcommand{\half}{\frac{1}{2}} 

\renewcommand{\bm}{\boldmath}

\def\be{\begin{equation}}
\def\ee{\end{equation}}
\def\bea{\begin{eqnarray}}
\def\eea{\end{eqnarray}}
\def\bes{\begin{subequations}}
\def\ees{\end{subequations}}

\renewcommand{\a}{\alpha}

\allowdisplaybreaks

\usepackage[square,numbers,compress]{natbib}
\usepackage{bibentry}

\counterwithin*{equation}{section}

\usepackage[hidelinks, linkcolor=blue, citecolor=blue, urlcolor=blue]{hyperref}

\begin{document}

\thispagestyle{empty}

\begin{flushright}
{IFT-UAM/CSIC-23-38}\\
\end{flushright}
\vskip .8 cm
\begin{center}
  {\Large {\bf\bm Symmetric fluxes and small tadpoles}}\\[12pt]

\bigskip
\bigskip 
\bigskip
{\bf Thibaut Coudarchet,}\footnote{thibaut.coudarchet@uam.es}\textsuperscript{,*}
{\bf Fernando Marchesano,}\footnote{fernando.marchesano@csic.es}\textsuperscript{,*}
{\bf David Prieto}\footnote{david.prietor@estudiante.uam.es}\textsuperscript{,*}\\
{\bf and Mikel A. Urkiola}\footnote{mikel.alvarezu@ehu.eus}\textsuperscript{,$\dagger$,$\ddagger$}

\setcounter{footnote}{0}
%To avoid starting the main text with a footnote number of 5.

\bigskip
\bigskip
\vspace{0.23cm}
\textsuperscript{*}{\it Instituto de F\'isica Te\'orica UAM-CSIC, c/ Nicol\'as Cabrera 13-15,\\ 28049 Madrid, Spain}\\[5pt] 
\textsuperscript{$\dagger$}{\it Department of Applied Mathematics, University of the Basque Country UPV/EHU, Plaza Ingeniero Torres Quevedo 1, 48013 Bilbao, Spain}\\[5pt]
\textsuperscript{$\ddagger$}{\it EHU Quantum Center, University of the Basque Country UPV/EHU, Bilbao, Spain}\\[20pt] 
\bigskip
\end{center}

\date{\today{}}

\begin{abstract}
\noindent

\vspace{10pt}

The analysis of type IIB flux vacua on warped Calabi--Yau orientifolds becomes considerably involved for a large number of complex structure fields. We however show that, for a quadratic flux superpotential, one can devise simplifying schemes which effectively reduce the large number of equations down to a few. This can be achieved by imposing the vanishing of certain flux quanta in the large complex structure regime, and then choosing the remaining quanta  to respect the symmetries of the underlying prepotential. One can then implement an algorithm to find large families of flux vacua with a fixed flux tadpole, independently of the number of fields. We illustrate this approach in a Calabi--Yau manifold with 51 complex structure moduli, where several reduction schemes can be implemented in order to explicitly solve the vacuum equations for that sector. Our findings display a flux-tadpole-to-stabilized-moduli ratio that is marginally above the bound proposed by the Tadpole Conjecture, and we discuss several effects that would take us below such a bound. 

\end{abstract}

\newpage 
\setcounter{page}{2}

\renewcommand{\baselinestretch}{1.5}

%%%%%%%%%%%%%%%%%%%%%%%%%%%%%%%%%%%%%%%%%%%%%%%%%%%

\tableofcontents

\section{Introduction}
\label{sec:Introduction}

Our perception of the string Landscape is conditioned by how well we understand the set of vacua that compose it and, in particular, the equations that describe such vacua. In this sense, type IIB or F-theory flux compactifications on warped Calabi--Yau (CY) orientifolds seem like a particularly simple setup. The energy-minimization constraints that background fluxes impose on the axio-dilaton/complex structure  vacuum expectation values (vevs) can be characterized in terms of a tree-level K\"ahler potential and superpotential \cite{Gukov:1999ya,Giddings:2001yu} (see \cite{Grana:2005jc,Douglas:2006es,Blumenhagen:2006ci,Becker:2006dvp,Marchesano:2007de,Denef:2008wq,Denef:2007pq,Ibanez:2012zz,Quevedo:2014xia,Baumann:2014nda} for reviews), and the corresponding flux vacuum conditions lead to a system with the same number of equations as of unknowns.\footnote{In this paper we do not discuss Kähler moduli stabilization, and the term {\em flux vacua} refers to those axio-dilaton/complex structure vevs that solve the equations of motion at tree-level in 4d Minkowski.} Algebraically, this points towards a discrete set of solutions, which is indexed by the set of  flux quanta. Since the number of flux quanta grows with the number of complex structure deformations, this reasoning would indicate that this sector of the string Landscape is dominated by CY manifolds with the largest number of such moduli. 

The problem of this naive picture is that such fluxes induce D3-brane charge and tension, measured by the integer $\Nf$, and this quantity cannot overshoot the negative charge and tension induced by the geometry of the compactification, or else flux vacua cannot be found. This additional constraint could in principle modify the above perception that the discretum of type IIB/F-theory flux vacua should be dominated by Calabi--Yau manifolds with a large complex structure sector. In fact, it was argued in \cite{Bena:2020xrh} that quite the opposite should be true, as a consequence of the Tadpole Conjecture (TC). This proposal essentially states that if $n_{\rm stab} \gtrsim {\cal O} (100)$ is a large number of flux-stabilized moduli, then the ratio
\begin{equation}
\Nf / n_{\rm stab}\ ,
\label{ratio}
\end{equation}
is bounded from below by $1/3$. 

The difficulty in testing this proposal is that, in general, the set of vacuum equations becomes intractable for a very large number of moduli. As a consequence, many of its tests are of numerical nature \cite{Bena:2020xrh,Bena:2021wyr,Tsagkaris:2022apo}, while analytic techniques to address the validity of the proposal are still being developed. Along these lines, \cite{Marchesano:2021gyv} provided a set of algebraic equations that describe smooth F-theory and type IIB flux vacua in the large complex structure (LCS) regime, for an arbitrary number of moduli. Using them, one can see that the value of $\Nf$ goes to infinity along some directions in field space, unless certain flux quanta are set to zero. Guided by this fact, \cite{Marchesano:2021gyv} proposed a set of  patterns for vanishing flux quanta which, in the case of type IIB compactifications were dubbed {\it IIB1} and {\it IIB2 scenarios}. While in the second of these setups $\Nf$ is independent of $n_{\rm stab}$,  which goes against the TC statement, it was argued in \cite{Plauschinn:2021hkp,Lust:2021xds,Grimm:2021ckh} that demanding control over exponentially-suppressed corrections could restore a linear dependence, in line with the TC proposal. Using asymptotic Hodge structure techniques, this intuition was made precise for strict asymptotic regimes in complex structure moduli space \cite{Grana:2022dfw}, finding a lower bound for the ratio \eqref{ratio} twice above the proposed value. Such strict asymptotic regimes probe  asymptotic limits along many directions in field space, which points towards a universal behaviour of the ratio \eqref{ratio} as we approach them. However, it is important to stress that they demand a strict hierarchy between the complex structure vevs, and as such they do not cover the whole of the LCS region. This leaves open the possibility that the ratio \eqref{ratio} lowers its value at other regions within the LCS regime, as well as on the deep interior of the complex structure moduli space. In fact, it was shown in \cite{Lust:2022mhk} that at the vicinity of Fermat points one can measure the above ratio to be two orders of magnitude below the initially proposed value.

Regardless of the validity of the Tadpole Conjecture and/or its refinements, evaluating the quantity \eqref{ratio} in different regions of fields space is interesting per se, as it measures the efficiency of this moduli stabilization mechanism in them. Indeed, at flux vacua $\Nf$ measures the energy stored in the flux background, and so the inverse of \eqref{ratio} can be understood as the number of stabilized moduli per flux-energy unit. It is thus a pressing question to determine if this moduli-stabilization  efficiency grows towards the interior of the moduli space, as the above set of results seem to indicate, since this would have a direct impact on our perception of the string Landscape. The problem is that the dataset at the deep interior of moduli space is very scarce, as the type of analysis made in \cite{Lust:2022mhk} (see also \cite{Braun:2020jrx,Becker:2022hse}) is only valid along invariant loci of discrete symmetries acting on the CY complex structure moduli space; loci that are selected upon choosing fluxes invariant under such a symmetry \cite{Giryavets:2003vd,Giryavets:2004zr,DeWolfe:2004ns,Denef:2004dm,Louis:2012nb,Cicoli:2013cha,Blanco-Pillado:2020wjn}. 

In this paper we aim to bridge the gap between the analysis made in \cite{Grana:2022dfw} and \cite{Lust:2022mhk}, by analysing flux moduli stabilization in a complex structure region in between these two regimes, as illustrated by figure \ref{fig:conjecture}. We consider flux vacua in the LCS region, so that we can make use of mirror symmetry techniques to write down equations that describe such vacua. Even though we are in an asymptotic regime of the CY moduli space, we focus on regions that are far away from the growth sectors considered in \cite{Grana:2022dfw}, which require a strict hierarchy between vevs. More precisely, we consider flux vacua in which many complex structure vevs are similar, which allows us to reduce the complexity of the system of equations down to a few of them. This selection is implemented by a choice of flux quanta that partially respects the symmetries of an underlying ${\cal N} =2$ prepotential, and in this sense our approach is similar in spirit to that of \cite{Lust:2022mhk}. However, we do not require a discrete moduli-space symmetry to implement our analysis,\footnote{Any such symmetry can only be approximate, as it is presumably broken by exponential effects.} and we do not need to integrate any field out of the 4d supergravity description. We dub this approach as an {\em effective reduction} of the system of vacua equations, as it is closer to the  ansatz-based philosophy implemented in \cite{Blanco-Pillado:2020hbw}.

More precisely, we develop our analysis based on the type IIB1 scenario of \cite{Marchesano:2021gyv}, whose analytical features were further analyzed and developed in \cite{Coudarchet:2022fcl}. It is thanks to this analytic control that we are able to get an overall picture of this setup, and propose a general algorithm to generate large families of flux vacua in the LCS region. Unlike in \cite{Coudarchet:2022fcl}, we do not focus on choices of flux quanta that correspond to the no-scale aligned vacua of \cite{Blanco-Pillado:2020wjn}. Indeed, as already pointed out in \cite{Marchesano:2021gyv}, this flux choice yields a ratio \eqref{ratio} strictly larger than one, and so it is not easy to implement for CY compactifications with $h^{2,1} \gg 1$. Nevertheless, there are other choices of flux quanta within the type IIB1 scenario that yield a much smaller ratio even for ${\cal O} (100)$ flux-stabilized moduli, as we will demonstrate with explicit examples. 

Our constructions are based on the  deformed $T^6/(\Z_2\times\Z_2)$ orbifold analyzed in \cite{Vafa:1994rv}, which hosts $51$ complex structure moduli, and which we dub as $(3,51)$ Calabi--Yau. An orientifold of this Calabi--Yau has been considered before in literature in order to find flux vacua \cite{Blumenhagen:2003vr,Cascales:2003zp}, but only at the orbifold limit. Unlike in \cite{Blumenhagen:2003vr,Cascales:2003zp} we are able to build consistent flux vacua that do not overshoot the D3-brane tadpole, because we allow for non-vanishing fluxes in the would-be orbifold twisted sector, that moreover stabilize the corresponding moduli in the LCS regime. This geometry will not only allow us to implement the IIB1 scenario, but also to do so via different reduction ans\"atze. We explore in particular reductions with two, four and six effective parameters, which constitute a huge simplification compared to the initial system of 104 real equations. In this setup, our algorithm allows to generate large families of flux vacua in the interior of the saxionic complex structure (CS) cone, and deep enough to be able to neglect exponential corrections, by varying a set of flux quanta that do not contribute to the tadpole. This allows us to provide counterexamples to some of the statements made in the literature based on asymptotic limits, like the expectation that $\Nf$ should increase as vacua go deeper into the interior of the CS cone. 

With these results in hand, we evaluate the ratio \eqref{ratio} in our constructions and find that, as expected, they lie in between the values found in \cite{Grana:2022dfw} and \cite{Lust:2022mhk}. Remarkably, the smallest of such values marginally respects the lower bound $1/3$ proposed in \cite{Bena:2020xrh}, meaning that the stabilization of an additional real modulus would violate it. In our case, this result is due to two separate effects. First, we do not include D7-brane position moduli in the counting of $n_{\rm stab}$, even if three-form fluxes generically stabilize such moduli \cite{Gorlich:2004qm,Camara:2004jj,Gomis:2005wc}. Second, we only consider orientifold geometries that contain standard O3-planes, which leads to flux quantization conditions demanding even flux quanta \cite{Frey:2002hf}. Dropping any of these two assumptions could easily lead to reducing the value of the ratio \eqref{ratio} by a factor of 2 or 4, violating the proposed bound of $1/3$. More generally, these remarks open the question on whether the tadpole-to-stabilized moduli ratio is sensitive to the presence of exotic O3-planes in a Calabi--Yau orientifold compactification, which so far is a rather unexplored corner of the string Landscape. 

The paper is organized as follows: in sect.~2, we briefly recap some notations and conventions for type IIB flux compactifications in the LCS regime. In sect.~3, we review the IIB1 flux scenario of \cite{Marchesano:2021gyv} with special emphasis on the properties derived from the quadratic  superpotential, which provides a simplifying scheme for moduli axio-dilaton and complex structure moduli stabilization. We also provide our definition of \emph{effective reduction} on the number of moduli, and see under which assumptions such reduction can occur in the IIB1 scenario. In sect.~4, we gather information about the mirror dual of the compactification geometry that we investigate in this work, namely the symmetrically resolved $T^6/(\Z_2\times \Z_2)$ orbifold with $h^{1,1}=51$. The topological data of this model is useful to express the LCS prepotential on the mirror side. In sect.~5, starting from the IIB1 vacuum equations for the compactification space at hand, we apply a reduction ansatz to effectively reduce the number of complex structure moduli down to two. We then search for vacua under the constraint that exponential corrections are negligible so as to justify the LCS approximation, and that the D3-brane tadpole bound is not overshot. In sect.~6, a new family of solutions is uncovered in a more complex four-parameter reduction of the model. Sect.~7 discusses how our results relate to the tadpole conjecture, and in sect.~8 we provide our conclusions and outlooks.  Several details are relegated to the appendices: appendix \ref{sec:Solutions} provides full numerical details about some vacua presented in the paper, while appendix \ref{sec:six-param} makes explicit the six-parameter reduction mentioned earlier.

\section{Type IIB flux compactifications}
\label{sec:notations}

In this section we review the material  used to describe the effective behaviour of flux compactifications of type IIB string theory on a Calabi--Yau 3-fold $X_3$ at large complex structure (LCS). In order to construct type IIB flux vacua one needs to introduce orientifold planes with negative D3-brane charge \cite{Dasgupta:1999ss,Giddings:2001yu}, which can be implemented via the standard O3/O7 projection based on a geometric involution ${\cal R}$,  such that ${\cal R} :\Omega \to - \Omega$ and ${\cal R} : J \to J$. Nevertheless, the search for flux vacua can be simply formulated in terms of the complex structure prepotential ${\cal F}$ for the Calabi--Yau covering space $X_3$, if one assumes that all harmonic three-forms are odd under the action of ${\cal R}$. In this section and the next we will take such a simplifying assumption, since the specific Calabi--Yau that we will analyze in the following sections satisfies this property. Our conventions and notation will follow those of \cite{Douglas:2006es} and \cite{Coudarchet:2022fcl}.\footnote{In particular, in our conventions $G_3 \wedge * \bar{G}_3 = \frac{1}{3!} G_3 \cdot \bar{G}_3 {\rm vol}_{X_6}$, where ${\rm vol}_{X_6} \equiv \frac{i}{8} \Omega \wedge \bar{\Omega} = \frac{1}{3!} J^3$. Note the sign differences with respect to the conventions in \cite{Giddings:2001yu}.}

\subsection{The complex structure prepotential}

In the LCS regime that will be of interest to us in this paper, the complex structure prepotential $\F$ for a Calabi-Yau $X_3$ takes the form
\begin{equation}
\F\equiv-\frac{1}{6}\kappa_{ijk}z^iz^jz^k-\half a_{ij}z^iz^j+c_iz^i+\half\kappa_0+\F_{\rm inst}\ ,
\label{eq:full_prepotential}
\end{equation}
where the complex structure fields are written $z^i$, $i=1,\dots,h^{2,1}(X_3)$. The instanton contribution $\F_{\rm inst}$ is exponentially suppressed in the LCS regime while the polynomial coefficients $\kappa_{ijk}$, $a_{ij}$, $c_i$ and $\kappa_0$ arise from topological data of the mirror manifold $Y_3$ of the Calabi--Yau $X_3$. These topological quantities are defined as follows \cite{Mayr:2000as}
\begin{align}
\begin{split}
\label{eq:kappa0}
&\kappa_{ijk}\equiv\int_{Y_3}\omega_i\wedge\omega_j\wedge\omega_k\ ,\qquad a_{ij}\equiv-\half\int_{Y_3}\omega_i\wedge i_*\text{ch}_1(\text{P.D}[w_j])\ ,\\
&c_i\equiv\frac{1}{24}\int_{Y_3}\omega_i\wedge \text{ch}_2(Y_3)\ ,\qquad  \kappa_0\equiv\frac{\zeta(3)\chi(Y_3)}{(2\pi i)^3}\ = i \, \frac{\zeta (3)}{4\pi^3} (h^{1,1}(Y_3) - h^{2,1}(Y_3))\ ,
\end{split}
\end{align}
where the 2-forms $\omega_i$, $i=1,\dots,h^{1,1}(Y_3)$ form a basis of $H^2(Y_3,\Z)$,  P.D stands for  Poincaré Dual, $i_*$  for the pushforward of the embedding $i$ of the divisors into the mirror $Y_3$ and eventually, $\text{ch}_1$ and $\text{ch}_2$ denote the first and second Chern classes respectively. The quadratic coefficient $a_{ij}$ can be expressed from the triple intersection numbers \cite{Cicoli:2013cha} through the relation
\begin{equation}
    a_{ij} = -\frac{1}{2} \int_{Y_3} \omega_i \wedge \omega_j \wedge \omega_j\ .
\end{equation}

The coefficients $c_i$ and $a_{ij}$ are defined only modulo $\mathbb{Z}$ as a consequence of redundancies when describing the complex structure moduli through a basis of 3-cycles of $X_3$. Thanks to these redundancies, we will use the equivalent convention of \cite{Tsagkaris:2022apo,Demirtas:2020ffz} to write the quadractic coefficient of the prepotential as
\begin{align}
    a_{ij} = - \frac{1}{2} \left\lbrace
    \begin{array}{ll}
        \kappa_{iij}, & i \geq j \\
        \kappa_{ijj}, & i < j
    \end{array}.
    \right.
\label{eq:aij}
\end{align}

From a symplectic basis of 3-cycles $\{A^I,B_I\}$, $I=0,\dots,h^{2,1}$ of $H_3(X_3,\Z)$, one can then express the periods of the Calabi--Yau $(3,0)$-form $\Omega$ like
\begin{equation}
\Pi^t\equiv\left(\int_{B_I}\Omega,\int_{A^I}\Omega\right)=(\F_I,X^I)\ ,
\end{equation}
where we have $z^i\equiv X^i/X^0$, $i=1,\dots,h^{2,1}$ and $\F_I$ denotes the derivative of the prepotential with respect to $X^I$. In the gauge $X^0=1$, the period vector then reads
\begin{equation}
\label{eq:period}
    \Pi = 
    \begin{pmatrix}
     2\mathcal{F} - z^i \partial_i \F \\
    \partial_i \F \\
     1 \\
    z^i 
    \end{pmatrix}.
\end{equation}

\subsection{The Kähler potential and superpotential}

At tree-level, the Kähler potential takes the form
\begin{eqnarray}\nonumber
K\equiv K_{\text{k}} + K_{\text{dil}} + K_{\text{cs}} & = & -2\log(\V)-\log(-i(\tau-\bar\tau))- \log\left(i\int_{X_3} \Omega \wedge \bar{\Omega} \right) \\
& =& -2\log(\V)-\log(-i(\tau-\bar\tau))-\log(-i\Pi^\dagger\cdot\Sigma\cdot\Pi)\ ,
\end{eqnarray}
where $\V$ stands for the volume of the Calabi--Yau $X_3$, $\tau$ is the axio-dilaton and the symplectic $(2h^{2,1}+2)\times(2h^{2,1}+2)$ matrix reads
\begin{equation}
\Sigma\equiv\begin{pmatrix}
0 & \mathds{1}\\ -\mathds{1} & 0
\end{pmatrix}.
\end{equation}
In the LCS regime, the complex structure part of the potential involving the period vector is given by
\begin{align}
    K_{\text{cs}} &= - \log \left( \frac{i}{6} \kappa_{ijk} (z^i-\bar{z}^i) (z^j-\bar{z}^j) (z^k-\bar{z}^k) - 2 \, \Im (\kappa_0) \right) \nonumber \\
    &= - \log \left( \frac{4}{3} \kappa_{ijk} t^i t^j t^k - 2 \, \Im (\kappa_0) \right)\ ,
\label{eq: complex Kahler potential}
\end{align}
where the complex structure fields can be  decomposed into axionic and saxionic parts: $z^i \equiv b^i + i t^i$. In a similar way, we define $\tau \equiv b^0 + i t^0$. \\

Fluxes threading the compact geometry induce a superpotential $W$, known as the Gukov-Vafa-Witten (GVW) superpotential \cite{Gukov:1999ya}. In order to describe it, we can introduce the following notation for the flux quanta
\begin{equation}
N\equiv f-\tau h\ \ \ \ \text{with}\ \ \ \ 
f\equiv\begin{pmatrix}
\int_{B^I} F_3 \\ 
\int_{A_I} F_3
\end{pmatrix} 
\equiv
\begin{pmatrix}
f^B_0 \\ f^B_i \\ f_A^0 \\ f_A^i 
\end{pmatrix}
\ \ \ \ \text{and}\ \ \ \
h\equiv\begin{pmatrix}
\int_{B^I}H_3\\
\int_{A_I}H_3
\end{pmatrix}
\equiv
\begin{pmatrix}
h^B_0 \\ h^B_i \\ h_A^0 \\ h_A^i 
\end{pmatrix} ,
\end{equation}
which leads to the following compact formulation of the flux superpotential\footnote{Note that we deliberately drop out an overall factor $1/\sqrt{4\pi}$.} \cite{Gukov:1999ya}
\begin{equation}
W\equiv   \int_{X_3} G_3 \wedge \Omega  \equiv \int_{X_3} (F_3-\tau H_3)\wedge \Omega = N^T \cdot \Sigma \cdot \Pi\ .
\end{equation}
This expression can be easily expanded in the LCS regime to yield
\begin{align}
\begin{split}
W = &- \frac{1}{6} N_A^0 \kappa_{ijk} z^i z^j z^k + \frac{1}{2} \kappa_{ijk} N_A^i z^j z^k + \left( N_A^j a_{ij} + N_i^B - N_A^0 c_i \right) z^i\\
&- \kappa_0 N_A^0 - N_A^i c_i + N_0^B\ .
\label{eq:Wfull}
\end{split}
\end{align}

Finally, it must be noted that fluxes induce a D3-tadpole Ramond-Ramond charge in the compact manifold, which must be cancelled by negatively charged objects, like orientifold planes. It can then  be shown that the full covering-space D3-charge $\Nf$ induced by these fluxes is
\begin{equation}
\label{eq:Nflux}
    \Nf = \int_{X_3} F_3 \wedge H_3 =  f^T \cdot \Sigma \cdot h  \ .
\end{equation}

\subsection{The vacuum equations}

The supergravity scalar potential famously enjoys a no-scale property in the Kähler sector such that the $4d$ vacua are in Minkowski spacetime. The vacuum equations are then simply expressed by requiring the covariant derivatives of the superpotential with respect to the axio-dilaton and complex structure fields to vanish: $D_I W\equiv\partial_IW+(\partial_IK)W= 0, \ I\in\left\lbrace \tau,z^i \right\rbrace$. More explicitly, making use of the definitions above, these equations read
\begin{align}
D_\tau W &= \left[ - h  - \frac{1}{\tau - \bar{\tau}} (f - \tau h) \right]^T \cdot \Sigma \cdot \Pi = - \frac{1}{\tau - \bar{\tau}} \bar{N}^T \cdot \Sigma \cdot \Pi = 0\ , \\[5pt]
D_i W &= N^T \cdot \Sigma \cdot D_i \Pi = 0\ .
\end{align}

%%%%%%%%%%%%%%%%%%%%%%%%%%%%%%%%%%%%%%%%%%%%%%%%%%%%%%%%%%%%%%%%

\section{Flux families and simplifying schemes}
\label{s:schemes}

In this section we briefly review the vacuum equations arising from a flux family which renders the superpotential quadratic in the axio-dilaton and complex structure moduli. This flux setup, dubbed \emph{IIB1 scenario}, has been defined in \cite{Marchesano:2021gyv} and extensively investigated in \cite{Coudarchet:2022fcl}. We also specify the notion of \emph{effective reductions} in the moduli space, while emphasizing how it is different from proper truncations \cite{Giryavets:2003vd,Giryavets:2004zr,DeWolfe:2004ns,Denef:2004dm,Louis:2012nb,Cicoli:2013cha,Blanco-Pillado:2020wjn,Kachru:2003aw,Balasubramanian:2005zx,Conlon:2005ki,
Westphal:2006tn,Klemm:1992tx,Doran:2007jw,Candelas:2017ive,Braun:2011hd,Batyrev:2008rp,Doran:2005gu,Candelas:2019llw,Joshi:2019nzi,Grimm:2019ixq}.

\subsection{Quadratic superpotentials at LCS}
\label{sec:IIB1}

The so-called \emph{IIB1 setup} \cite{Marchesano:2021gyv,Coudarchet:2022fcl} has been shown to simplify the search for solutions of no-scale vacua. It is defined by the following flux constraints:
\begin{equation}
\label{eq:def_IIB1}
\text{IIB1 flux configuration:}\quad f_A^0=0\ ,\  h_A^0=0\ \text{ and }\ h_A^i=0\ ,\ i\in\{1,\dots,h^{2,1}\}\ ,
\end{equation}
from where it is easy to check that the flux superpotential $W$ in \eqref{eq:Wfull} is quadratic at any point on moduli space on the  axio-dilaton and complex structure moduli. In particular, it takes the form
\begin{equation}
W=\half\vec{Z}^t M \vec{Z} + \vec{L} \cdot \vec{Z} +Q\ ,
\label{eq:W_bilinear}
\end{equation}
where $\vec{Z}\equiv (\tau, \vec{z})$ and where the $(h^{2,1}+1)$-dimensional matrix $M$, the vector $\vec L$ and the scalar $Q$ are real flux-dependent quantities:
\begin{equation}
\label{eq:MLQ}
M\equiv\begin{pmatrix}
0 & -\vec h^{B\, t}\\
-\vec h^{B} & S_{ij}
\end{pmatrix},
\quad \vec L\equiv(-h_0^B,f_i^B+a_{ij}f_A^j)\ ,\quad Q\equiv f_0^B-c_if_A^i\ ,
\end{equation}
with $S_{ij} \equiv \kappa_{ijk} f_A^k$. The vanishing fluxes also simplify greatly the expression of the flux-induced D3-brane charge $\Nf$, which reads
\begin{align}
N_{\text{flux}} = - f_A^i h_i^B .
\label{eq:nflux}
\end{align}
Therefore, in this setup, one is able to tune the remaining flux quanta $f_0^B$, $h_0^B$ and $f_i^B$ to one's liking, without this choice affecting the tadpole.

\subsubsection{Vacuum equations}
\label{sec:vacuum_equations}

The IIB1 choice of fluxes has many perks when considering the associated vacuum equations. One nice feature is a decoupling between the axionic and saxionic equations.

\paragraph{Axions:}

On the one hand, it is shown that the axions, i.e., the fields $b^0 \equiv \Re (\tau)$ and $b^i \equiv \Re (z^i)$ gathered into the vector $\vec B\equiv (b^0,b^i)$, obey a very simple linear equation:
\begin{equation}
\label{eq:MB}
M\vec B=-\vec L\ .   
\end{equation}

When we assume the matrix $S$ to be invertible with the further requirement that $\cH\equiv h_i^B S^{ij} h_j^B\neq 0$, then the matrix $M$ has maximal rank and we can invert the relation \eqref{eq:MB}. Moreover, the invertibility of $S$ allows to compute an analytical ``block-inverse'' for $M$ from which we deduce the vevs
\begin{align}
\label{eq:bs}
b^0 = \frac{h_i^B S^{ij} L_j - h_0^B}{h_i^B S^{ij} h_j^B }\ , \qquad\qquad
b^i = S^{ij} \left( b^0 h_j^B - L_j \right) \ .
\end{align}
Note that symplectic transformations on the fluxes will yield transformations on these fields due to the monodromy symmetry of the period vector at LCS. These transformations will act on the $b^A$, $A\in\{0,i\}$ as $b^A \to b^A + 1$, hence their name.

When the matrix $M$ is singular, the linear relation \eqref{eq:MB} only stabilizes $r\equiv\rank (M)$ axions and $h^{2,1}+1-\rank (M)$ constraints on the flux quanta arise.  This can be easily seen as follows \cite{Coudarchet:2022fcl}: we can diagonalize the matrix $M$ to $D\equiv\text{diag}(\lambda_0,\dots,\lambda_{r-1},0,\dots,0)$ with $\lambda_0,\dots,\lambda_{r-1}$ representing the $r$ non-zero eigenvalues of the matrix, and where there are as many zeroes as the dimension of the kernel. Denoting $N$ the change-of-basis matrix, we have
\begin{equation}
M= N^tDN\quad\text{ with }\quad N^t=N^{-1}\ .
\end{equation}
Defining $\vec B'\equiv N\vec B$ and $\vec L'\equiv N\vec L$, the axionic system of equations \eqref{eq:MB} becomes
\begin{equation}
D\vec B'=-\vec L'\ .
\end{equation}
Splitting the $h^{2,1}+1$ indices $\{0,i\}$ like $\alpha\in\{0,\dots,r-1\}$ and $\beta\in\{r,\dots,h^{2,1}\}$, the axionic vacuum expectation values and the flux constraints are given by
\begin{equation}
\label{eq:flux_constraints}
b'^\alpha=-\frac{\vec L'^\alpha}{\lambda_\alpha}\quad\text{ and }\quad \vec L'^\beta=0\ .
\end{equation}

\paragraph{Saxions:}

The saxionic system composed of $t^0 \equiv \Im (\tau)$ and $t^i \equiv \Im (z^i)$ does not generically enjoy such analytical solutions unless an additional ansatz is imposed \cite{Coudarchet:2022fcl}. The general vacuum equations for these fields read
\begin{align}
\left\lbrace
\begin{array}{l}
e^{-K_{\rm cs}}\left( S_{ij} t^j - t^0h_i^B \right) + 4 t^0 \kappa_{ijk}t^jt^k\left[h_l^Bt^l\right] =0\ , \\[10pt] 
\dfrac{1}{2} S_{ij} t^i t^j + t^0 h_i^B t^i = Q'\ ,
\end{array}
\right.
\label{eq:saxion_syst}
\end{align}
where we have defined the flux-dependent quantity
 \begin{equation}
\label{eq:Qp_singular}
Q'\equiv f_0^B-f_A^i c_i -\half \vec{L}^t M^+ \vec{L} \ ,
\end{equation}
with $M^+$ denoting the generalized inverse of $M$. It is defined like $M^+\equiv N^t D^+ N$ where $D^+ \equiv\text{diag}(\lambda_0^{-1},\dots,\lambda_{r-1}^{-1},0,\dots,0)$ and $\lambda_i$, $i\in\{0,\dots,r-1\}$ represent the non-zero eigenvalues of $M$. When $M$ is regular, $M^+=M^{-1}$ and we can write   
\begin{align}
Q' \equiv f_0^B - f_A^i c_i + \frac{(h_i^B S^{ij} L_j - h_0^B)^2}{2 h_i^B S^{ij} h_j^B} - \frac{1}{2} L_i S^{ij} L_j\ .
\label{eq:Qp}
\end{align}
Note that even though this system  generically requires the use of some numerical method to be solved, it is quite an improvement from the more generic vacuum equations, since half of the real variables (the axions) follow directly from \eqref{eq:MB}. Indeed, this system is composed of $h^{2,1} + 1$ equations and variables, so its numerical solution is expected to be obtained in a far more reduced time than in a more generic flux setup. 
The following constraint can be easily derived if both equations in \eqref{eq:saxion_syst} are combined:
\begin{align}
    Q' = - 3 t^0 \Im (\kappa_0) e^{K_{cs}} h_i^B t^i\ .
    \label{eq:Q_constraint}
\end{align}
We will make use of this relation in the following sections, as its numerical application will be quite useful.

As a final comment, it is interesting to note that the full system \eqref{eq:saxion_syst} can be reworked into a more manageable form by introducing the following decomposition:
\begin{align}
t^i \equiv t^0 v^i , \quad v^i \in \mathbb{R} \ . \label{eq:ri}
\end{align}
Indeed, it is straightforward to check that after this change of variables \eqref{eq:saxion_syst} becomes
\begin{align}
&\left[ \frac{4}{3} \kappa_{klm} v^k v^l v^m - \frac{2 \Im (\kappa_0)}{(t^0)^3} \right] \left( S_{ij} v^j - h_i^B \right) +  4 \kappa_{ikl} v^k v^l \left[ h_m^B v^m \right] = 0 \ ,\label{eq:saxion_syst_ri} \\[15pt] 
& t^0 = \sqrt{\dfrac{2Q'}{S_{ij} v^i v^j + 2 h_i^B v^i}} \ .\label{eq:saxion_syst_t0}
\end{align}
We should remark that the decomposition \eqref{eq:ri} is not an ansatz per se for the saxion fields $t^i$ and is simply a redefinition. However, it will become quite useful for the numerical search of vacua that we describe in the following subsections.

\subsubsection{System of equations close to the LCS point}

One of the nice things to consider about the system \eqref{eq:saxion_syst_ri}--\eqref{eq:saxion_syst_t0} is that it greatly simplifies  close to the LCS point, where
\begin{equation}
\label{eq:xi}
e^{-K_{\rm cs}}=\frac{4}{3} \kappa_{ijk} t^i t^j t^k - 2 \, \Im (\kappa_0) \approx \frac{4}{3} \kappa_{ijk} t^i t^j t^k\ \Longleftrightarrow\ |\xi|\equiv\left|\frac{-3\Im(\kappa_0)}{2\kappa_{ijk} v^i v^j v^k (t^0)^3}\right|\ll 1\ .
\end{equation}
The parameter $\xi$ is called the \emph{LCS parameter} since it indicates proximity with the LCS point, as long as in addition, all saxions have large vevs. It is easy to check that in that regime the system of equations becomes
\begin{align}
&\kappa_{klm} v^k v^l v^m \left( S_{ij} v^j - h_i^B \right) + 3 \kappa_{ikl} v^k v^l \left[ h_m^B v^m \right] = 0 \ , \label{eq:saxion_syst_LCS} \\[10pt] 
&(t^0)^2 \left[ \frac{1}{2} S_{ij} v^i v^j + h_i^B v^i \right] = Q' \ . \label{eq:t0_LCS}
\end{align} 

It is quite interesting to see that \eqref{eq:saxion_syst_LCS} completely determines all the $v^i$ in terms of only $f_A^i$ and $h_i^B$, i.e., the fluxes which contribute to the D3-tadpole. Therefore, one can numerically solve \eqref{eq:saxion_syst_LCS} in terms of this reduced set of fluxes, depending on how large we want the tadpole of the system to be. On the other hand, it is easy to check that the term in brackets in eq.~\eqref{eq:t0_LCS} is identically zero whenever \eqref{eq:saxion_syst_LCS} is satisfied. As such, $t^0$ is left as a free parameter of the system and this imposes some conditions upon the fluxes which come into play within $Q'$. Then, as long as one is close to the LCS point, the $v^i$ found with \eqref{eq:saxion_syst_LCS} will be a solid first-order approximation in order to solve the full system \eqref{eq:saxion_syst}.

\subsubsection{Strategy to solve the saxionic system numerically}
\label{sec:algorithm}

Taking all of the discussion above into consideration, we propose the following strategy to get a full solution of the saxion system \eqref{eq:saxion_syst} without, in principle, imposing any further ansatz on the solution: 

\begin{enumerate}
\item Find tuples of $\lbrace f_A^i, h_i^B \rbrace$ which satisfy the tadpole condition: $0<N_{\text{flux}} \leq L_\ast$ where $L_*$ is some upper bound not to be overshot. Using these, solve \eqref{eq:saxion_syst_LCS} in order to get a zeroth-order approximation to the $v^i$, written $v^i_{(0)}$.

\item Assume a target value for the dilaton, which we  dub $t^0_{\rm target}$. This initial value must be such that it corresponds to a region of LCS flux vacua. In practice, this means that $t^0_{\rm target} \in (t^0_{\rm min}, t^0_{\rm max})$, where the bounds are determined as follows:

\begin{itemize}

\item[-]  $t^0_{\rm min}$ is such that $t^i_{(0)} \equiv v^i_{(0)} t^0_{\rm min}$ is at the boundary of  the stretched complex structure (CS) cone \cite{Demirtas:2018akl,Plauschinn:2021hkp}. For non-simplicial cones like in the example that we will analyze, this condition can be quite non-trivial. 

\item[-] $t^0_{\rm max}$ is such that  the value for $Q'$ associated to $t^0_{\rm target}$, expressed via  \eqref{eq:Q_constraint} and \eqref{eq: complex Kahler potential} as
\begin{equation}
\label{eq:Qptarget}
Q'_{\rm target} \equiv \frac{-3\Im(\kappa_0)h_i^Bv^i_{(0)}(t^0_{\rm target})^2}{\frac{4}{3}\kappa_{ijk}v^i_{(0)}v^j_{(0)}v^k_{(0)}(t^0_{\rm target})^3-2\Im(\kappa_0)}\ ,
\end{equation}
reaches the value $Q'_{\rm min}$, defined as the smallest fraction in absolute value\footnote{At large $t^0$, the sign of $Q'$ is determined by that of $\Im(\kappa_0)$. In the model under study, where we have few Kähler moduli and a large number of complex structure fields, $Q'$ will be negative.} that one can get from expression \eqref{eq:Qp_singular}. Note that $Q'_{\rm min}$ depends on the fluxes contributing to $\Nf$, which have been fixed in step 1. Since asymptotically $Q'_{\rm target} \sim \pm 1/t^0_{\rm target}$, a minimal absolute value value for $Q'_{\rm target}$ translates into an upper bound $t^0_{\rm max}$ for $t^0_{\rm target}$.\footnote{Note that to determine $Q'_{\rm target}$, we make use of the zeroth-order values $v^i_{(0)}$ but in reality, they are good approximates of the actual ratios $t^i/t^0$ in the full solutions that we will build only when the LCS parameter is sufficiently small, i.e. for a sufficiently large dilaton vev. However in practice, this requirement is much less stringent than the lower bound $t^0_{\rm min}$ defined above.}

\end{itemize}

\item Find the remaining fluxes $\lbrace f_0^B, h_0^B, f_i^B \rbrace$ such that $Q'$ expressed with \eqref{eq:Qp_singular} or \eqref{eq:Qp} is as close as possible to the value $Q'_{\rm target}$ found above.

\item Solve the full system for the fluxes $f_A^i$, $h_i^B$ found in step 1 and the remaining fluxes determined in step 3. If $t^0_{\rm target}$ has been appropriately chosen sufficiently large and if $Q'$ is indeed tuned close to the targeted value, the exact solution will feature a final $t^0$ very close to $t^0_{\rm target}$ and $v^i$ close to the zeroth-order approximation of step 1.
\end{enumerate}

One may wonder why we should take all of these extra steps, instead of just directly solving the system \eqref{eq:saxion_syst_ri}-\eqref{eq:saxion_syst_t0} for certain flux tuples. This becomes self-evident when trying to solve the system by brute force. Indeed, if one wants to make a full search along the flux lattice and look for solutions over many different tuples of flux quanta (by, for example, taking random values for fluxes), one will eventually stumble upon the realization that very few fluxes allow for either solutions with a reasonable tadpole, or within the CS cone. The algorithm presented above (which is only applicable to the IIB1 family) allows for an efficient search for solutions within the CS cone, due to the decomposition \eqref{eq:ri}, all the while reducing the search for vacua with flux tuples which are explicitly within the tadpole bound as a first step.

\subsection{Effective reductions vs. truncations of moduli}

As mentioned above, the split between axions and saxions in the vacuum equations in the IIB1 scenario is a nice improvement compared to what a generic choice of fluxes would yield. However, the system of equations \eqref{eq:saxion_syst_ri} and \eqref{eq:saxion_syst_t0} involving the saxions is still $(h^{2,1}+1)$-dimensional, which becomes costly to solve numerically for large $h^{2,1}$. This is where \emph{effective reductions} can help, by reducing the number of variables and equations to solve.

An effective reduction is characterized by an ansatz on the moduli as well as on the various fluxes involved in the vacuum equations. The reduction is successful if one can obtain a reduced number of $n$ equations involving $n$ degrees of freedom. As an example, in \cite{Blanco-Pillado:2020hbw} a type IIB flux configuration was proposed together with an ansatz on the moduli to effectively reduce any model at LCS down to one modulus. 
 Note that what we dub an \emph{effective reduction} here is different from the  \emph{consistent supersymmetric truncations} that are abundantly used and studied in the literature \cite{Giryavets:2003vd,Giryavets:2004zr,DeWolfe:2004ns,Denef:2004dm,Louis:2012nb,Cicoli:2013cha,Blanco-Pillado:2020wjn}.  In the latter case, the truncated moduli are frozen at some fixed point (usually $0$) of a discrete symmetry action on the moduli space of the geometry under consideration. The truncated moduli disappear from the effective supergravity action and since they are deep in the interior of moduli space, getting information about their behaviour is a complicated task \cite{Lust:2022mhk}. In the case of an effective reduction, the flux setup along with the ansatz on the moduli fields only simplify the vacuum equations into those of an effective $n$-parameter model. In other words, the reduction ansatz specifies a submanifold of the moduli space on which the search for vacua is restricted. A proper flux choice then ensures that solutions of the reduced theory automatically give solutions to the full theory.

In the specific case of the IIB1 system of equations \eqref{eq:saxion_syst_ri}, it is easy to see that simplifications can occur if the triple intersection numbers and the fluxes enjoy simple symmmetry properties. More concretely, the whole equation depends on the flux $h_i^B$ and on contractions of the intersection numbers which schematically look like $\kappa_{ijk} A^j B^k$, where $A^i$ represents either the flux $f_A^i$ or the real variable $v^i$. \emph{Therefore, given some $v^i$ living in some $n$-dimensional subspace of the whole $\mathbb{R}^{h^{2,1}}$ (i.e., written in terms of only $n$ variables), as long as the covectors $\kappa_{ijk} f_A^j v^k$, $\kappa_{ijk} v^j v^k$ and $h_i^B$ have the same structure of repeated entries, such that only $n$ of them are independent, we will be able to reduce \eqref{eq:saxion_syst_ri} to $n$ equations.}

The purpose of the remainder of the paper is to illustrate this idea in a specific setup. In particular, we review the properties of a specific compactification Calabi--Yau, write the prepotential at LCS for this geometry, apply the IIB1 scheme and eventually propose effective reduction ans\"atze to find solutions with all axio-dilaton and complex structure moduli stabilized in the LCS regime.

\section{\bm A Calabi--Yau with 51 complex structure moduli}
\label{sec:orb_51}

In this section we review the key features of the symmetric resolution of a $T^6/(\Z_2\times\Z_2)$ orbifold model with $(h^{1,1}, h^{2,1}) = (51,3)$, which we dub $(51,3)$ Calabi--Yau. Upon mirror symmetry, this manifold maps into a deformation of the $T^6/(\Z_2\times\Z_2)$ orbifold with opposite choice of discrete torsion and 51 complex structure deformations \cite{Vafa:1994rv}, which we will dub as the $(3,51)$ Calabi--Yau. It is on an orientifold of this second manifold where we want to study the IIB1 scenario and  implement different effective reductions. 

The symmetric resolution of  $T^6/(\Z_2\times\Z_2)$ with 51 K\"ahler moduli has been extensively studied in the literature  \cite{Denef:2005mm,Faraggi:2021uli} among other toroidal orbifolds \cite{Reffert:2006du,Lust:2006zh}, so that all topological data of interest is available. Thanks to this information, we are able to express the relevant Kähler cone and stretched Kähler cone conditions of this manifold, that suffers from a severe non-simpliciality. This allows us to write the prepotential for the complex structure sector of the mirror dual $(3,51)$ Calabi--Yau, and check whether we are in the LCS region of moduli space where exponentially suppressed corrections can be neglected. As we will see, the symmetries displayed by the triple intersection numbers of the $(51,3)$ Calabi--Yau allow us to define well-behaved reduction ans\"atze for the IIB1 scenario in the $(3,51)$ Calabi--Yau.

\subsection{Topological data of the mirror}
\label{ss:topomirror}

We gather here the relevant topological numbers for the K\"ahler sector of the $(51,3)$ Calabi--Yau, useful to write the prepotential of the vector multiplet sector either for type IIA string theory compactified in this manifold or for type IIB compactified in the mirror, on which we then focus. We extract the information we need from the references \cite{Reffert:2006du,Lust:2006zh}, that investigate resolved orbifolds and \cite{Denef:2005mm,Faraggi:2021uli}, that explore this specific $\mathbb{Z}_2\times  \mathbb{Z}_2$ model. In our case, we will focus on the symmetric resolution.  The orbifold group is given by two generators $\{g_1,g_2\}$, whose action over the complex compact coordinates $\{z_i\}$ is 
\begin{align}
        g_1: \quad & z_{2,3}\rightarrow -z_{2,3}  & g_2:  \quad & z_{1,3}\rightarrow -z_{1,3} & g_1 \circ g_2: \quad & z_{1,2} \rightarrow -z_{1,2}
    \\
            & z_1\rightarrow z_1  &   & z_2\rightarrow z_2  & & z_3\rightarrow z_3 \ . \nonumber
\end{align}
As a consequence of the toroidal identification, there are $16$ fixed lines under the action of each group element ($48$ in total). These lines intersect along $64$ fixed points in which the blow up is performed. Therefore, the resulting orbifold has $h^{1,1}=51$ (the original $3$ untwisted Kähler moduli plus the $48$ twisted Kähler moduli associated to the blow up of the fixed lines) and $h^{2,1}=3$ (the complex structure is completely determined by providing the modular parameters of the three $T^2$ factors). A suitable basis for the divisors is given by
\begin{equation}
\{R_1,R_2,R_3\}\cup\{E_{i\alpha,j_{\beta}},\ i,j=1,2,3,\ i<j,\ \alpha,\beta=1,\dots,4\}\ .
\end{equation}
 The $E_{i\alpha,j\beta}$ are the exceptional divisors that arise from the resolution of the orbifold and are associated to the twisted sector. The $R_i$ are the sliding divisors descending from the unresolved model and  can be expressed in terms of the exceptional divisors and the toric divisors $D_{i\alpha}$ of the local model:
\begin{align}
\begin{split}
\label{Rs}
&R_1=2D_{1\alpha}+\sum_{\beta}E_{1\alpha,2\beta}+\sum_{\gamma}E_{3\gamma,1\alpha}\qquad \forall\alpha\ ,\\
&R_2=2D_{2\beta}+\sum_{\alpha}E_{1\alpha,2\beta}+\sum_{\gamma}E_{2\beta,3\gamma}\qquad \forall\beta\ ,\\
&R_3=2D_{3\gamma}+\sum_{\beta}E_{2\beta,3\gamma}+\sum_{\alpha}E_{3\gamma,1\alpha}\qquad \forall\gamma\ .
\end{split}
\end{align}
With this basis, the Kähler form $J$ is parametrized as
\begin{equation}
\label{eq:J}
J\equiv r_iR_i-t_{1\alpha,2\beta}E_{1\alpha,2\beta}-t_{2\beta,3\gamma}E_{2\beta,3\gamma}-t_{3\gamma,1\alpha}E_{3\gamma,1\alpha}\ .
\end{equation}

Notice that the conventions for labelling the divisors differ in the literature. We find useful to make explicit the dictionary between the different notations as follows:
\begin{equation}
\begin{aligned}
\label{eq:dictionary}
\text{\cite{Denef:2005mm}}\qquad&\longleftrightarrow\qquad \text{\cite{Faraggi:2021uli,Reffert:2006du,Lust:2006zh}} \\
E_{1\alpha,2\beta} \qquad&\longleftrightarrow\qquad  E_{3,\alpha\beta} \\
E_{2\beta,3\gamma} \qquad&\longleftrightarrow\qquad  E_{1,\beta\gamma} \\
E_{3\gamma,1\alpha} \qquad&\longleftrightarrow\qquad  E_{2,\alpha\gamma}
\end{aligned}
\end{equation}

\subsubsection{Triple intersection numbers}

Theses numbers are useful to express the cubic and quadratic terms in the prepotential \eqref{eq:full_prepotential} thanks to \eqref{eq:aij}. The non-zero intersection numbers are
\begin{equation}
\begin{aligned}
&R_1\cdot R_2\cdot R_3=2\ ,\qquad\quad &&E_{1\alpha,2\beta}\cdot E_{2\beta,3\gamma}\cdot E_{3\gamma,1\alpha}=1\ ,\quad &&\\
&R_1\cdot E_{2\beta,3\gamma}^2=-2\ ,&&R_2\cdot E_{3\gamma,1\alpha}^2=-2\ , &&R_3\cdot E_{1\alpha,2\beta}^2=-2\ ,\\
&E_{2\beta,3\gamma}^2\cdot E_{3\gamma,1\alpha}=-1\ ,&&E_{2\beta,3\gamma}\cdot E_{3\gamma,1\alpha}^2=-1\ ,&&E_{2\beta,3\gamma}^2\cdot E_{1\alpha,2\beta}=-1\ ,\\
&E_{2\beta,3\gamma}\cdot E_{1\alpha,2\beta}^2=-1\ ,&&E_{3\gamma,1\alpha}^2\cdot E_{1\alpha,2\beta}=-1\ ,&&E_{3\gamma,1\alpha}\cdot E_{1\alpha,2\beta}^2=-1\ ,\\
&E_{1\alpha,2\beta}^3=4\ , &&E_{2\beta,3\gamma}^3=4\ ,&&E_{3\gamma,1\alpha}^3=4\ .
\end{aligned}
\label{eq:div_intersections}
\end{equation}
Since it will be useful in the next paragraph, we also write the self triple intersection of the divisors $D_{i\alpha}$ which is $D_{i\alpha}^3=8$.

\subsubsection{Intersection with second Chern class}

The intersections of the divisors with the second Chern class ${\rm ch}_2(Y_3)$ are useful to express the linear term in the prepotential. To compute them, we make use of the following formula \cite{Reffert:2006du,Lust:2006zh}
\begin{equation}
{\rm ch}_2(Y_3)\cdot P+P^3=\chi(P)\ ,
\end{equation}
where $P$ stands for any divisor and $\chi(P)$ is its Euler characteristic. Finding the intersections with ${\rm ch}_2(Y_3)$ thus amounts to knowing the topology of the divisors. From \cite{Denef:2005mm}, we read that the topology of the toric divisors $D_{i\alpha}$ is that of $\mathbb{P}^1\times \mathbb{P}^1$ while the topology of the exceptional divisors is that of $\mathbb{P}^1\times\mathbb{P}^1$ blown-up in $4$ points (we denote this $\text{Bl}_4(\mathbb{P}^1\times\mathbb{P}^1)$). We thus have
\begin{align}
\begin{split}
\label{eq:chE}
&{\rm ch}_2(Y_3)\cdot D_{i\alpha}=\chi(\mathbb{P}^1\times\mathbb{P}^1)-D_{i\alpha}^3=4-8=-4\ ,\\
&{\rm ch}_2(Y_3)\cdot E_{i\alpha,j\beta}=\chi(\text{Bl}_4(\mathbb{P}^1\times\mathbb{P}^1))-E_{i\alpha,j\beta}^3=8-4=4\ .
\end{split}
\end{align}
From this along with the relations (\ref{Rs}) and by schematically denoting any exceptional divisor like $E$, we deduce
\begin{equation}
\label{eq:chR}
{\rm ch}_2(Y_3)\cdot R_i=2c_2(Y_3)\cdot D_{i\alpha}+8c_2(Y_3)\cdot E=-8+32=24\ .
\end{equation}
Note that according to \cite{Reffert:2006du}, such an intersection with ${\rm ch}_2(Y_3)$ for the $R_i$ is associated to a $K3$ topology.

\subsection{The prepotential at LCS}

We now derive the prepotential describing the complex structure sector of the $(3,51)$ Calabi--Yau in the large complex structure regime, mirror to the resolved orbifold described above in the large volume regime. As explained for instance in \cite{Cicoli:2013cha}, we need to identify the (complexified) Kähler moduli of the $(51,3)$ CY with the complex structure moduli of the $(3,51)$ CY. In terms of the Kähler parameters introduced above, we apply the following prescription:
\begin{equation}
\begin{aligned}
&z^i,\ i=1,2,3 \to r_i\ , &&i=1,2,3\ ,\\
&z^i,\ i=4,\dots,19 \to t_{2\beta,3\gamma}\ ,\qquad\qquad &&\gamma,\alpha=1,\dots,4\ ,\\
&z^i,\ i=20,\dots,35 \to t_{3\gamma,1\alpha}\ ,\ &&\alpha,\beta=1,\dots,4\ ,\\
&z^i,\ i=36,\dots,51 \to t_{1\alpha,2\beta}\ ,\ &&\beta,\gamma=1,\dots,4\ ,
\end{aligned} 
\end{equation}
where we order the pairs of indices $(\beta,\gamma)$, $(\gamma,\alpha)$ and $(\alpha,\beta)$ in lexicographical order: $(1,1)$, $(1,2)$, $(1,3)$, $(1,4)$, $(2,1)$, $\dots$

We begin with the cubic term of the LCS prepotential. As discussed in \cite{Candelas:1990pi,Hosono:1994av}, the so-called Yukawa couplings of the cubic terms of the prepotential, i.e., the $\kappa_{ijk}$ factors, have to match the ones obtained when writing the volume of the Calabi--Yau as
\begin{align}
\V = \frac{1}{6} \kappa_{ijk} s^i s^j s^k,
\end{align}
where $s^i = \lbrace r_i , t_i \rbrace$. This means that the triple intersection numbers can be read off from \eqref{eq:div_intersections}, taking into account that odd powers of $E$ divisors will carry an extra negative sign within their respective intersections. On the other hand, the polynomial corrections to be added in the LCS prepotential can be computed using the expressions given in section \ref{sec:notations}. The quadractic corrections are expressed with \eqref{eq:aij} from the triple intersection numbers. For the linear coefficients $c_i$, we use the intersections of the divisors with the second Chern class \eqref{eq:chE} and \eqref{eq:chR} to find
\begin{equation}
c_i=(1,1,1,1/6,\dots,1/6)\ .
\end{equation}
The constant term is determined by the Euler characteristic of the Calabi--Yau, see \eqref{eq:kappa0}. For this particular example, we find
\begin{equation}
\kappa_0=\frac{12i\zeta(3)}{\pi^3}\ .
\end{equation}

\subsection{Kähler and Mori cones of the mirror}
\label{sec:kahler_mori}

The Kähler cone of a Calabi--Yau compactification is defined as the collection of Kähler forms $J$ such that all curves, areas and volumes computed with respect to it are positive%, i.e., \cite{Plauschinn:2021hkp} 
\begin{align}
\label{eq:stretched}
\mathcal{K}_Y \equiv \left\lbrace J \in H^{1,1} (Y, \mathbb{R}) \ | \ \text{vol}(W)>0 \ \forall W \in \mathcal{W} \right\rbrace ,
\end{align}
where $\mathcal{W}$ includes all curves, areas and volumes constructible with $J$. In practice, we will be mostly interested in exploring a somewhat more constrained version of this cone, namely the \emph{stretched} Kähler cone \cite{Demirtas:2018akl}, where all of the volumes are required to satisfy $\text{vol}(W)>c$ with $c=\mathcal{O}(1)$, in order to be free from potential exponential corrections. As in  \cite{Plauschinn:2021hkp}, we will dub its mirror as stretched complex structure (CS) cone.

In the symmetric resolution of the blown-up $T^6/(\mathbb{Z}^2 \times \mathbb{Z}^2)$, the volumes to consider are the following \cite{Faraggi:2021uli} (we translate notations according to \eqref{eq:dictionary}):
\begin{align}
&\text{vol}(R_1 R_2) = 2 r_3\ , \label{eq:vol_first}\\[4pt]
&\text{vol}(R_1 E_{2\beta,3\gamma}) = 2 t_{2\beta,3\gamma}\ , \\[4pt]
&\text{vol}(R_1 D_{2\beta}) = r_3 - \sum_{\gamma} t_{2\beta,3\gamma}\ , \label{eq:vol_b1}\\[4pt]
&\text{vol}(D_{1\alpha} E_{3\gamma,1\alpha}) = r_2 - \sum_\beta t_{1\alpha,2\beta}\ ,\label{eq:vol_b2} \\[4pt]
&\text{vol}(E_{2\beta,3\gamma}E_{3\gamma,1\alpha}) = t_{2\beta,3\gamma} + t_{3\gamma1\alpha} - t_{1\alpha,2\beta}\ ,\label{eq:vol_curves} \\[4pt]
&\text{vol}(R_1) = 2 r_2 r_3 - \sum_{\beta, \gamma} t_{2\beta,3\gamma}^2\ ,\\[4pt]
&\text{vol}(D_{1,\alpha}) = r_2 r_3 - \sum_{\gamma} r_2 t_{3\gamma,1\alpha} - \sum_{\beta} r_3 t_{1\alpha,2\beta} + \sum_{\beta,\gamma} t_{3\gamma,1\alpha} t_{1\alpha,2\beta}\ , \\[4pt]
&\text{vol}(E_{2\beta,3\gamma}) = 2 r_1 t_{2\beta,3\gamma} + \sum_{\alpha} \bigg[ \frac{1}{2} t_{2\beta,3\gamma}^2 + t_{3\gamma,1\alpha} t_{1\alpha,2\beta} - \frac{1}{2} (t_{3\gamma,1\alpha}^2+ t_{1\alpha,2\beta}^2 )\\[4pt]
&\hspace{8.5cm}- t_{2\beta,3\gamma} t_{3\gamma,1\alpha} - t_{2\beta,3\gamma} t_{1\alpha,2\beta} \bigg]\nonumber\ , \\[4pt]
&\text{vol}(Y_3) = 2 r_1 r_2 r_3 - \sum_{\beta,\gamma} r_1 t_{2\beta,3\gamma}^2 - \sum_{\alpha,\gamma} r_2 t_{3\gamma,1\alpha}^2 - \sum_{\alpha \beta} r_3 t_{1\alpha,2\beta}^2  \ , \label{eq:vol_last}\\[4pt]
&-\sum_{\alpha,\beta,\gamma} \left[ - \frac{1}{2} t_{2\beta,3\gamma} (t_{3\gamma,1\alpha}^2 + t_{1\alpha,2\beta}^2) - \text{perms} + \frac{1}{6} \left(  t_{2\beta,3\gamma}^3 + t_{3\gamma,1\alpha}^3 +  t_{1\alpha,2\beta}^3 \right) + t_{2\beta,3\gamma} t_{3\gamma,1\alpha} t_{1\alpha,2\beta} \right]\nonumber\ ,
\end{align}
and their corresponding cyclic permutations.

The Mori cone is the cone spanned by all effective curves. These curves are characterized by their intersections with the divisors of the basis, that we can read from the formulas \eqref{eq:vol_first}-\eqref{eq:vol_curves} and arrange into $51$-dimensional vectors. For example, from \eqref{eq:vol_first}, we read that the curve $R_1\cdot R_2$ is described by the vector $(0,0,1,0,\dots,0)$. From \eqref{eq:vol_curves}, we read that the curve $E_{2\beta,3\gamma}\cdot E_{3\gamma,1\alpha}$ with $\alpha=\beta=\gamma=1$ is described by the vector
\begin{equation}
(0,0,0,1,\underbrace{0,\dots,0}_{\text{15 times}},1,\underbrace{0,\dots,0}_{\text{15 times}},-1,\underbrace{0,\dots,0}_{\text{15 times}})\ .
\end{equation}
In total, this gives $267$ curves that generate the Mori cone. However, a \emph{basis of generators} is enough. Meaning, we can remove all curves that can be expressed as linear combinations with positive coefficients of the others. From \cite{Denef:2005mm}, the relevant basis of generators\footnote{We ran the algorithm described in \cite{Denef:2004dm} to explicitly build the basis of generators by removing positive linear combinations and we fully agree with the resulting set.} is given only by the curves \eqref{eq:vol_b1}, \eqref{eq:vol_b2} and \eqref{eq:vol_curves} and their permutations. This gives a basis of generators with $216$ elements. The Mori cone is thus highly non-simplicial in the sense that it is generated by many more elements than its dimension.

In order to be safe from exponential corrections, we require volumes of all subvarieties to be greater than $1$ in string units.
This means we want to be inside the stretched Kähler cone defined with a parameter $c=1$. From a mirror perspective, we want the complex structure moduli to be such that the expressions for the various volumes are greater than $1$ when replacing the real parts of the Kähler moduli by the saxionic parts of the mirror's complex structures.

\subsection{Orientifolding, flux quantization and the D3-brane tadpole}
\label{sec:quantization}

In order to construct type IIB flux vacua based on the $(3,51)$ Calabi--Yau $X_3$, we need to introduce orientifold planes with negative D3-brane charge in the above construction. We do so via the standard O3/O7 projection based on a geometric involution ${\cal R}$,  such that ${\cal R} :\Omega \to - \Omega$ and ${\cal R} : J \to J$. The fixed loci of this involution will host O3-planes and O7-planes, and we will assume that the Ramond-Ramond charge of the latter is cancelled by a set of D7-branes, possibly on top of the O7-planes. The remaining RR tadpole condition then reads
\begin{equation}
\Nf+2N_{\rm D3}+Q_{\rm D3}=0\ ,   
\label{tadpoleD3}
\end{equation}
where $\Nf > 0$ is the flux-induced contribution \eqref{eq:Nflux}, $N_{\rm D3}$ counts the number of D3-branes in the quotient space $X_3/{\cal R}$ and the D3 charge $Q_{\rm D3}$ contains contributions from the O3-planes as well as from the D7-branes and O7-planes. When this last charge is negative, it sets an upper bound on the value of $\Nf$, if one forbids anti D3-branes or other non-BPS objects. It follows that the allowed set of flux vacua depends on the value of $Q_{\rm D3}$, which in turn depends on the orientifold projection. 

Simple choices of orientifold projections are those that are defined at the toroidal orbifold limit $T^6/(\Z_2\times\Z_2)$. They are described by a set of discrete choices on the Chan-Paton degrees of freedom, that are compatible with the choice of discrete torsion that corresponds to the $(3,51)$ Calabi--Yau, see e.g. \cite{Klein:2000qw} for a discussion. Out of these choices, we are interested in those that lead to the type IIB orientifold considered in \cite{Cascales:2003zp}, and that is T-dual to the type I orientifold constructed in \cite{Berkooz:1996dw}. Just like in \cite{Cascales:2003zp} we will consider three-form fluxes on top of this orientifold background but, unlike \cite{Cascales:2003zp}, we will not restrict ourselves to flux vacua that occur at the orbifold limit. Instead, we will also consider vacua at points in complex structure moduli space in which the collapsed three-cycles of the orbifold gain a non-trivial size. To do so in the presence of the orientifold projection, it is important to realize that the full cohomology of twisted three-forms is odd under this orientifold projection, which means that the projection is compatible with the complex structure moduli space of the $(3,51)$ Calabi--Yau. 

Given this particular orientifold projection, the precise O-plane content at the orbifold limit depends on the choice of discrete B-field along the non-trivial two-cycles of the compactification  \cite{Cascales:2003zp}. In the absence of such a B-field, the content amounts to 64 O3$^-$-planes and three sets of 4 O7$^-$-planes, where the superindex indicates the sign of the charge and tension compared to the corresponding D-branes. One may then cancel the O7-plane charge by placing 4 D7-branes plus their orientifold images on top of each O7-plane,\footnote{Intersecting D7-brane configurations with worldvolume fluxes may yield vacua with semi-realistic chiral spectra \cite{Cascales:2003zp,Marchesano:2004yq,Marchesano:2004xz}, but such magnetic fluxes increase the value of $Q_{\rm D3}$. Therefore for simplicity we will not consider them in our setup.} leaving an O3-plane charge that can be cancelled by a combination of D3-branes and three-form fluxes. In the absence of fluxes, the number of D3-branes in the covering space that cancels the tadpole is 32, which means that in \eqref{tadpoleD3} we have
\begin{equation}
Q_{\rm D3}= - 32 \implies \Nf\leq 32\ .   
\label{TD3basic}
\end{equation}
By continuity, it is easy to argue that the same result will hold when this orientifold projection is extended to the full complex structure moduli space of the $(3,51)$ Calabi--Yau. Indeed, deforming the orbifold geometry by giving a non-trivial vev to the twisted fields should not change the number of D3-branes that are needed to cancel tadpoles. In particular it is not expected that the growth of collapsed three-cycles induces curvature terms on the 7-brane sector that could contribute to $Q_{\rm D3}$ in one way or the other. 

Notice that in \cite{Cascales:2003zp} no flux configuration was found with $N_{\rm flux} \leq 32$. The reason was the quantization conditions imposed for three-form fluxes in the orbifold limit, which required that $N_{\rm flux} \geq 64$. In our analysis we will not have to impose such a lower bound because, unlike in \cite{Cascales:2003zp}, we allow for three-form fluxes switched along the twisted three-cycles. Indeed, having the above prepotential, we are entitled to include such fluxes as long as the resulting vacua lie within the LCS regime of the $(3,51)$ Calabi--Yau manifold. In the following sections we will see that with these more relaxed constraints and using certain effective reduction ans\"atze, one is able to find flux configurations that stabilize most of the moduli and do not overshoot the D3-brane tadpole \eqref{TD3basic}. 

Despite our more general framework, there are still certain consistency conditions that must be observed. In particular, one needs to consider the flux quantization condition derived by Frey and Polchinski in \cite{Frey:2002hf}, which states that the three-form flux quanta along any three-cycle that intersects an even number of O3$^-$-planes must be even. In the setup described above, in which there is no discrete B-field and there are 64 O3$^-$-planes, this constraint will apply to any relevant three-cycle, as can be checked in the orbifold limit. Therefore both NSNS and RR three-form flux quanta must be even, which implies $N_{\rm flux} \in 4\Z$, and severely constrains the allowed set of flux vacua. Switching on a discrete B-field along one or several K\"ahler moduli will introduce O3$^+$-planes in the configuration, as discussed in \cite[section 6]{Cascales:2003zp}, and may imply that flux quanta along certain three-cycles must be odd. However, each of these non-vanishing discrete B-fields will halve the value of $Q_{\rm D3}$ compared to \eqref{TD3basic}, since O3$^+$-planes contribute with a positive charge and tension to the D3-brane tadpole. Thus, the simplest approach is to focus on finding flux vacua with even flux quanta and $N_{\rm flux} \leq 32$. 

Alternatively, one may consider orientifold projections different from the ones discussed above, and that are not necessarily compatible with the orbifold limit. An example of such an O3/O7 projection was given in \cite{Denef:2005mm}, for the smooth $(51,3)$ Calabi--Yau manifold discussed in subsection \ref{ss:topomirror}, mirror to the one of interest in this paper. In order to specify the value of $Q_{\rm D3}$ for these alternative orientifold projections one would first have to define a holomorphic involution ${\cal R}$ on the smooth $(3,51)$ Calabi--Yau, the fixed loci hosting O3 and O7-planes and their charges, and finally the topology of the four-cycles wrapped by the O7-planes. Alternatively, one could find the F-theory four-fold uplift of the orientifold and compute its Euler characteristic. Since both problems are quite involved in general, one may instead resort to a lower bound estimate for the D3 charge $Q_{\rm D3}$, widely used in the type IIB orientifold literature \cite{Tsagkaris:2022apo,Collinucci:2008pf,Carta:2020ohw,Crino:2022zjk,Gao:2022fdi}, relevant when the D7-brane tadpole is cancelled locally with $SO(8)$ stacks:
\begin{equation}
 \label{eq:QD3}
 Q_{\rm D3}\geq -(2+h^{1,1}+h^{2,1})\ .
\end{equation}
In our case, this bound translates into $Q_{\rm D3}\geq -56$ which in turn sets an upper bound on the flux-induced contribution
\begin{equation}
\label{eq:boundNf}
\Nf\leq 56\ .    
\end{equation}
Note that this value matches the one obtained in \cite{Denef:2005mm} for the $(51,3)$ Calabi--Yau manifold with the alternative orientifold projection defined therein. 

The search for flux vacua in a given Calabi--Yau orientifold,  like the procedure described in section \ref{s:schemes}, mostly depends on the prepotential for the complex structure sector, and only through the constraints mentioned above on the O-plane content of the compactification. Therefore one may run a general analysis in the search of flux vacua for the $(3,51)$ Calabi--Yau manifold independently of the orientifold projections, using \eqref{eq:boundNf} as a guideline for the allowed flux tadpole.  Since this more general approach does not specify the nature of the O3-planes of the compactification, one may consider both even and odd fluxes, knowing that in order to verify the validity of the flux vacuum one should embed these flux choices in a specific choice of O-plane content, like the one leading to the more stringent bound \eqref{TD3basic}. In the following sections we will adopt this general philosophy when discussing reduction schemes and their solutions, although special emphasis will be given to those flux vacua that are compatible with the orientifold projection leading to \eqref{TD3basic}.

\section{Reduction to two parameters and stabilization}
\label{sec:two-param}

Now that we have reviewed all the necessary data to define the LCS prepotential for the $(h^{1,1}, h^{2,1})=(3,51)$ Calabi--Yau, we will turn to solve the vacuum equations \eqref{eq:saxion_syst_ri}-\eqref{eq:saxion_syst_t0} for some particular setup which allows for an easy numerical treatment, following the strategy outlined in subsection \ref{sec:algorithm}.

\subsection{Reduction ansatz}

In order to solve the system of vacuum equations, we propose the following ansatz for saxions and fluxes, inspired by \cite{Denef:2005mm} and motivated by the symmetries of the triple intersection numbers:\footnote{Note that while in section \ref{sec:kahler_mori} $r$ and $t$ represent Kähler moduli of the mirror $(51,3)$ Calabi--Yau, here they represent complex structure moduli of the $(3,51)$ Calabi--Yau.}
\begin{align}
\begin{aligned}
t^i &\equiv t^0 v^i  \equiv (r,r,r,\underbrace{t,\dots,t}_{\text{48 times}})\ , \\[5pt] 
v^i &\equiv (v^r,v^r,v^r,\underbrace{v^t,\dots,v^t}_{\text{48 times}})\ , \\[5pt]
f_A^i &\equiv (f_A^r,f_A^r,f_A^r,\underbrace{f_A^t,\dots,f_A^t}_{\text{48 times}})\ , \\[5pt]
h_i^B &\equiv (h_r^B, h_r^B, h_r^B, \underbrace{h_t^B,\dots,h_t^B}_{\text{48 times}})\ . 
\end{aligned}
\label{eq:2_par_ansatz}
\end{align}
The remarkable feature about this particular choice of fluxes and vevs is that it reduces the system \eqref{eq:saxion_syst_ri} (and thus also the system \eqref{eq:saxion_syst_LCS} close to the LCS point) to only two  equations.\footnote{Meaning that in total, the 52 equations for the saxions are composed of 3 and 49 identical equations.}  Together with eq.~\eqref{eq:saxion_syst_t0}, the whole system reduces to the following three equations:
\begin{align}
0&=-h_r^B+4f_A^rv^r-32f_A^tv^t+\frac{24\left[h_r^Bv^r+16h_t^Bv^t\right]\left[\left(v^r\right)^2-8\left(v^t\right)^2\right]}{8\left[\left(v^r\right)^3-24v^r\left(v^t\right)^2+48\left(v^t\right)^3\right]-\frac{\Im(\kappa_0)}{\left(t^0\right)^3}}\ , \nonumber\\[7pt]
0&=-\left[h_t^B+2f_A^t(v^r - 6 v^t)+2f_A^rv^t\right]-\frac{24(v^r-3v^t)v^t(h_r^Bv^r+16h_t^Bv^t)}{8\left[\left(v^r\right)^3-24v^r\left(v^t\right)^2+48\left(v^t\right)^3\right]-\frac{\Im(\kappa_0)}{\left(t^0\right)^3}}\ ,\nonumber\\[7pt]
t^0&=\left(\frac{Q'}{3h_r^B v^r + 48 v^t \left[h_t^B - 2 f_A^t v^r + 6 f_A^t v^t\right] + 6 f_A^r \left[(v^r)^2 - 8 (v^t)^2\right]}\right)^{\half}\ .
\end{align}
The equations at the LCS point are obtained by taking $\Im(\kappa_0)=0$ in these expressions. Note that for this particular restricted choice of fluxes, the flux-induced D3-brane tadpole reads
\begin{align}
N_{\text{flux}} = -3 f_A^r h^B_r - 48 f_A^t h^B_t \ .
\end{align}

\subsection{Constraints on the solutions}

Inserting the ansatz \eqref{eq:2_par_ansatz} into the relevant volumes gives\footnote{Notice that there are some overall factors of $2$ compared to \cite{Denef:2005mm} because we are computing volumes on the Calabi--Yau covering space.} 
%no orientifold projection has been applied.}
\begin{equation}
\begin{aligned}
&\V_{Y_3}=2\left(r^3-24rt^2+48t^3\right)\ ,\quad &&\V_E=2\left(rt-3t^2\right)\ ,\quad &&\V_D=(r-4t)^2\ ,\\[5pt]
&\A_r=r-4t\ ,\quad &&\A_t=t\ .
\end{aligned}    
\end{equation}
These expressions can be understood in terms of the volumes and areas of the mirror $(51,3)$ Calabi--Yau. In particular, $\V_{Y_3}$ stands for the volume of the whole mirror compactification space, $\V_E$ and $V_D$ are respectively the volumes of its divisors $E_{i\alpha,j\beta}$ and $D_{i\alpha}$, and the areas $\A_r$ and $\A_t$ are obtained from eqs.~\eqref{eq:vol_b1}, \eqref{eq:vol_b2} and \eqref{eq:vol_curves}. In the $(3,51)$ Calabi--Yau, these quantities translate into the quotient of three-cycle volumes  by the reference three-cycle volume. 

The stretched CS cone conditions with a parameter $c\geq 1$ then take the simple form
\begin{equation}
c<t<\frac{r-c}{4}\ .    
\end{equation}
which, using the decomposition \eqref{eq:ri}, gives
\begin{align}
v^r > 0\ , \qquad 
\left\lbrace
\begin{array}{l}
0 < v^t < \frac{v^r}{5} \ , \quad t^0 > \frac{c}{v^t} \\[5pt]
\frac{v^r}{5} < v^t < \frac{v^r}{4} \ , \quad t^0 > \frac{c}{v^r - 4 v^t}
\end{array}
\right. .
\label{eq:r_conditions}
\end{align}
Again by using mirror symmetry, one can interpret these conditions as those sufficient for exponentially suppressed corrections to be negligible in the search for flux vacua.

\subsection{Search for vacua}

In the following subsection we will go over the steps described in sect.~\ref{sec:algorithm} to find potential tree-level flux vacua for the $(3,51)$ Calabi--Yau.

\subsubsection{Solving the LCS system}

\paragraph{Step 1:}  Guided by \eqref{eq:boundNf}, we generate tuples $\{f_A^r,f_A^t,h_r^B,h_t^B\}$ such that the flux-induced contribution to the tadpole is lower than $56$. Then, we try to solve the system of equations at LCS \eqref{eq:saxion_syst_LCS} to find $v^r_{(0)}$ and $v^t_{(0)}$ under the constraints \eqref{eq:r_conditions} displayed above.\footnote{Meaning that we impose $v^r_{(0)}>0$, $v^t_{(0)}>0$ and if $v^t_{(0)}>v^r_{(0)}/5$, we ask $v^r_{(0)}>4v^t_{(0)}$. These conditions ensure positive volumes that can then be tuned to the desired values in the following steps.} The solutions found with smallest tadpoles are summarized in table \ref{tab:two_parameter_LCS}. We indicate the rank of the matrix $M$, defined in \eqref{eq:MLQ}, %corresponding to the choices of flux.  We display 
as well the value of the product $(t^0)^3 |\xi|$ defined in \eqref{eq:xi}, which is a function of $v^r_{(0)}$ and $v^t_{(0)}$. Note that both $t^0$ and the LCS parameter $\xi$ are unfixed at this stage.

\begin{table}
\centering
\begin{tabular}{c|cccc|cc|c|c}
$N_{\text{flux}}$ & $f_A^r$ & $f_A^t$ & $h_r^B$ & $h_t^B$ & $v^r_{(0)}$ & $v^t_{(0)}$ & $\rank (M)$ & $(t^0)^3\, |\xi|$\\ \hline
36 & -6 & -2 & 2 & 0 & 0.589 & 0.110 & 52 & 0.599\\
48 & -8 & -2 & 2 & 0 & 0.227 & 0.029 & 41 & 7.07
\end{tabular}
\caption{Two numerical solutions found for the two-parameter reduction at LCS. Each solution is associated with another one featuring opposite fluxes, that we do not write for brevity. We also indicate the rank of the matrix $M$, that counts stabilized axions.}
\label{tab:two_parameter_LCS}
\end{table}

\subsubsection{\bm Full solutions}
\label{sec:full_sols}

Let us now detail the three next steps of the algorithm to get solutions of the full system from these two first-step candidates. Note that as reviewed in sect.~\ref{sec:vacuum_equations}, the rank of the matrix $M$ determines how many axionic directions are stabilized. When the rank is maximal, all axions get a mass and no constraint arise. When the matrix is singular, only $\rank (M)$ axions are stabilized and $52-\rank (M)$ constraints on the flux quanta arise in our setup. For the solution with $\Nf=48$ in table~\ref{tab:two_parameter_LCS}, this means that $11$ axions will remain massless and special care will be required in the following steps to account for the flux constraints mentioned above.

\paragraph{Step 2:} We want to fix the target value for $t^0$ in order to  get a small LCS parameter (e.g. $|\xi|\sim10^{-2}$) and satisfy the stretched CS cone conditions \eqref{eq:r_conditions} with parameter $c=1$. For the solutions at hand and the rather small associated values for $(t^0)^3\,|\xi|$, the stretched cone conditions are the most stringent. We thus choose $t^0_{\rm target}=1/v^t_{(0)}$  since $v^t_{(0)}<v^r_{(0)}/5$ for both solutions. This value corresponds to $t^0_{\rm min}$ defined in sect.~\ref{sec:algorithm} for a stretched cone with parameter $1$. With this choice, we evaluate the target value $Q'_{\rm target}$ from \eqref{eq:Qptarget}. At this step, one can choose to aim for a larger value of $t^0$, to have some room with the bounds or if one desires to generate solutions inside a more stretched CS cone, i.e. one defined with a larger parameter $c$. Actually, to account for the fact than when generating the remaining flux quanta, one may obtain a definitive $t^0$ a bit lower than the target, we increase its value by 15\% to have some room.
\begin{itemize}
    \item For the $\Nf=36$ solution, we set $t^0_{\rm target}=10.46$ and obtain $Q'_{\rm target}=-0.304$.
     \item For the $\Nf=48$ solution, we set $t^0_{\rm target}=39.32$ and obtain $Q'_{\rm target}=-0.367$.
\end{itemize}

\paragraph{Step 3:} We fix all the $53$ remaining fluxes $\{f_0^B,f_i^B,h_0^B\}$ by minimizing $(Q'-Q'_{\rm target})^2$ over the integers, with $Q'$ expressed in terms of fluxes with \eqref{eq:Qp} when $M$ is invertible and with \eqref{eq:Qp_singular} when it is not. In the singular case, we must additionally impose the constraints \eqref{eq:flux_constraints} on the flux quanta for the axionic vacuum equations to be consistent. The remaining free fluxes can be fixed efficiently in full generality without further restrictions. 
\begin{itemize}
    \item For our $\Nf=36$ example, a flux choice can be 
    \begin{align}
    (f_{1}^B,\dots,f_{51}^B,f_0^B,h_0^B)=&(4,0,-2,2,-2,-6,-2,0,2,-2,0,-2,10,-2,8,0,\nonumber\\
&8,-2,-2,2,0,2,-4,4,2,2,-2,0,0,-4,0,0,-4,4,\nonumber\\
&2,-2,-2,-2,2,-4,-2,0,0,-2,0,2,0,-4,0,2,-4,\nonumber\\
&4,-2)
    \label{eq:config36}
    \end{align}
    which yields a definitive value $Q'=-0.302$.
    \item For the $\Nf=48$ solution, one can easily implement the $11$ constraints and tune the remaining flux quanta to one's liking. A possible flux tuple is given by 
    \begin{align}
    (f_{1}^B,\dots,f_{51}^B,f_0^B,h_0^B)=&(0,0,0,0,-6,-4,-8,-12,-8,8,4,0,6,0,-4,-4,-2,\nonumber\\
&0,6,2,-2,-4,8,4,0,-2,-4,-6,-14,2,2,-2,0,0,-8,\nonumber\\
&0,-2,-2,-6,6,0,-2,0,-2,0,0,-6,2,-2,-10,0)
    \label{eq:config48}
    \end{align}
    which yields the final value $Q'=-0.375$.
\end{itemize}

\paragraph{Step 4:} We now plug all the flux quanta into the full system of equations and solve it numerically. The table \ref{tab:full_sols} summarizes the flux choices for the two solutions and displays the final vevs for the saxions, while table \ref{tab:volumes} shows the relevant volumes and areas of the unmirrored manifold at the vacua. For a more complete display of the solution with $\Nf=36$, refer to appendix \ref{sec:sol_36}.

\begin{table}[ht!]
\centering
\begin{tabular}{c|cccc|c|ccc}
$N_{\text{flux}}$ & $f_A^r$ & $f_A^t$ & $h_r^B$ & $h_t^B$ & $ (f_{1}^B,\dots,f_{51}^B,f_0^B,h_0^B)$ & $t^0$ & $r$ & $t$\\ \hline
36 & -6 & -2 & 2 & 0 & See eq.~\eqref{eq:config36} & 10.5 & 6.19 & 1.16\\
48 & -8 & -2 & 2 & 0 & See eq.~\eqref{eq:config48} & 38.5 & 8.74 & 1.13
\end{tabular}
\caption{Two full solutions with $\Nf=36$ and $\Nf=48$.}
\label{tab:full_sols}
\end{table}

\begin{table}[ht!]
\centering
\begin{tabular}{c|ccccc}
$\Nf$ & $\V_{Y_3}$ & $\V_{E}$ & $\V_{D}$ & $\A_r$ & $\A_t$ \\ \hline
36 & 226 & 6.30 & 2.45 & 1.57 & 1.16\\
48 & 939 & 12.07 & 17.91 & 4.23 & 1.13
\end{tabular}
\caption{The volumes and areas in the vacua displayed above.}
\label{tab:volumes}
\end{table}

\subsubsection{Full families of solutions}
\label{sec:families}

It is important to realize that the two flux vacua presented above are just specific examples of the full set of  solutions that one obtains by applying the strategy of subsection \ref{sec:algorithm}. Indeed, the content of table \ref{tab:two_parameter_LCS} is related to step 1 of our flux vacua search algorithm, and in our discussion of step 2 above we chose a specific value for $t^0_{\rm target}$ close to the lower bound $t^0_{\rm min}$. Recall however that there is a whole range of values that one can choose for $t^0_{\rm target}$ in step 2 of the algorithm. Varying this value will generate a whole family of flux vacua associated to the content of table 1, where the flux quanta that vary are those that do not contribute to the tadpole. In the following we describe how to generate this family of flux vacua for the two examples discussed above. 

\begin{figure}
    \centering
    \includegraphics[scale=0.48]{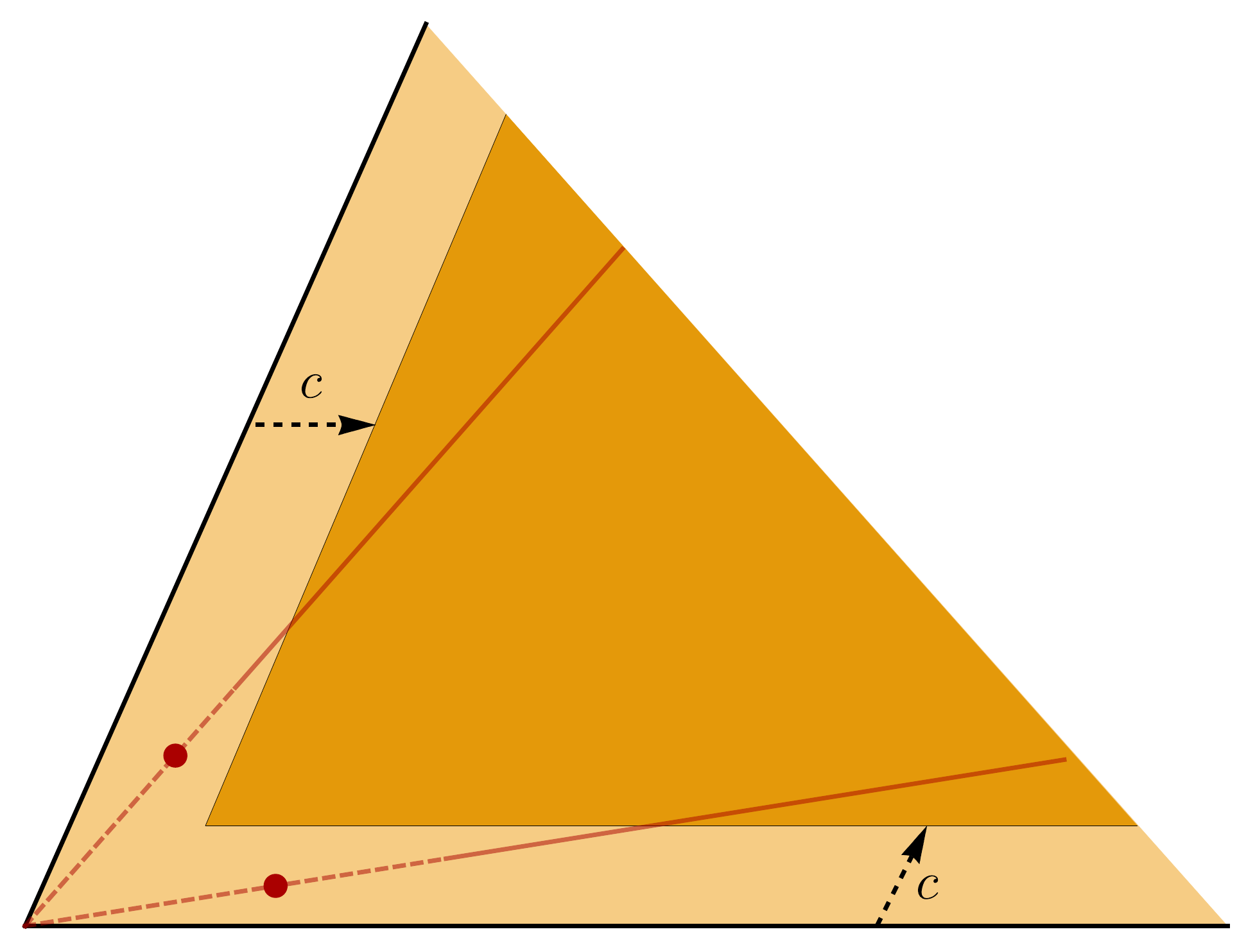}
    \caption{The CS cone and its strectched version with parameter $c$. The first-step results represented by the red dots define rays coming from the origin along which we have $t^i/t^0=v^i_{(0)}$. The design is inspired by \cite[figure 1]{Tsagkaris:2022apo}.}
    \label{fig:cone}
\end{figure}

To obtain the two full solutions above, the strategy was to aim at a sufficiently large $t^0_{\rm target}$ such that:
\begin{itemize}
\item We sit in the stretched CS cone.
\item We are deep enough in the LCS region such that the final ratios $t^i/t^0$ of the solution are very close to the zeroth-order values given by $v^i_{(0)}$.
\end{itemize}
We can generate further solutions based on the content of table \ref{tab:two_parameter_LCS}, either by going deeper inside the CS cone, or on the contrary closer to its boundary. To do this, one simply needs to repeat the whole procedure outlined above for the two specific solutions, but now scanning over different values for $t^0_{\rm target}$ in step 2. The CS cone depicted in figure \ref{fig:cone} illustrates this strategy: the red dots indicate the locations of the points in moduli space corresponding to saxions obtained from the first-step values $v^i_{(0)}$ with $t^0=1$. They are thus the points with coordinates $t^i=v^i_{(0)}$ and $t^0=1$. Note that they are in the CS cone by construction since we look for positive volumes in step 1, but in general they will not belong to the stretched CS cone. These points define rays starting from the origin of the cone, along which we have $t^i/t^0=v^i_{(0)}$. For increasing values of $t^0$, the lines go deeper inside the CS cone and eventually reach the stretched cone with parameter $c=1$. The first portions of the red lines are dashed to emphasize that $v^i_{(0)}$ is an accurate approximation of the values of the ratios in actual solutions only if $t^0$ is large enough, as reminded just above. True solutions coming out of our strategy are thus expected to deviate from the dashed part but converge towards the solid portion of the red lines.

\begin{figure}[t]
    \centering
    \begin{subfigure}[t]{0.49\textwidth}
    \centering
    \includegraphics[scale=0.21]{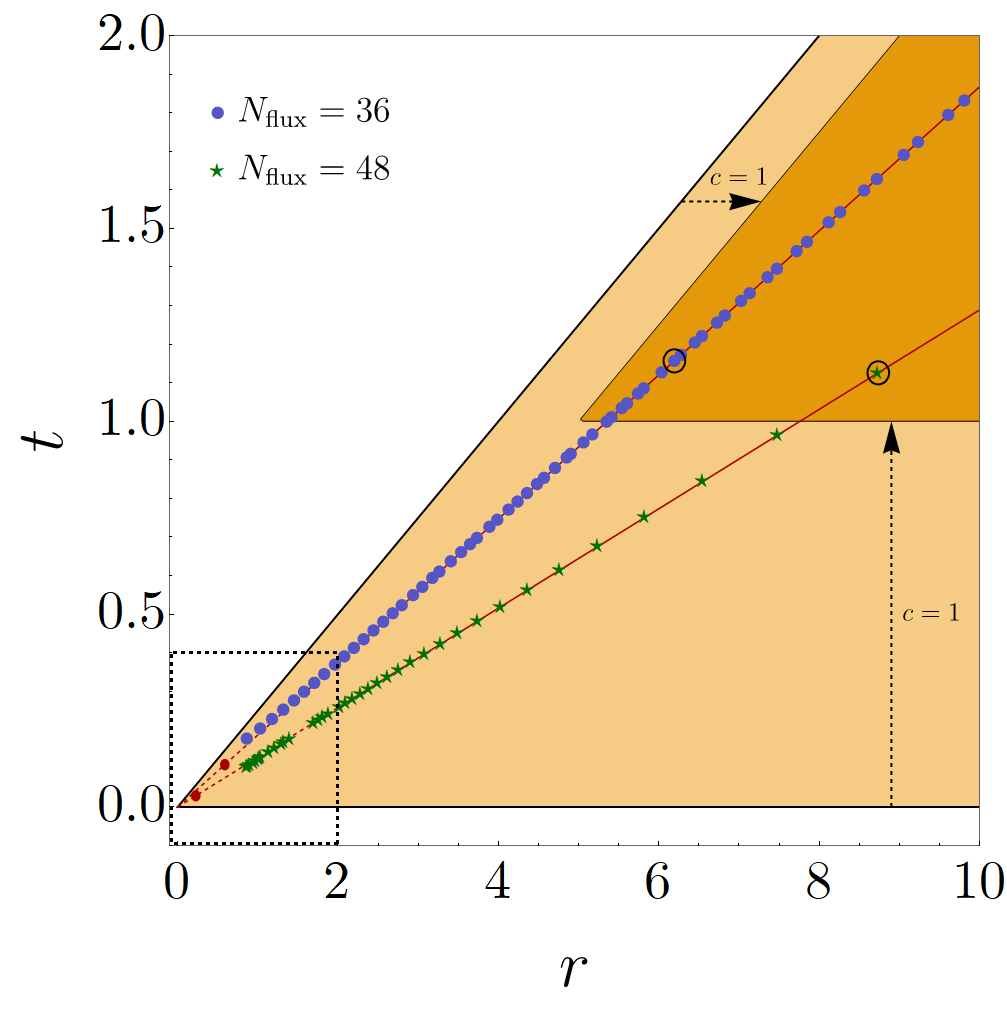}
    \label{fig:cone_true}
    \end{subfigure}
    \ 
    \begin{subfigure}[t]{0.49\textwidth}
    \centering
    \includegraphics[scale=0.211]{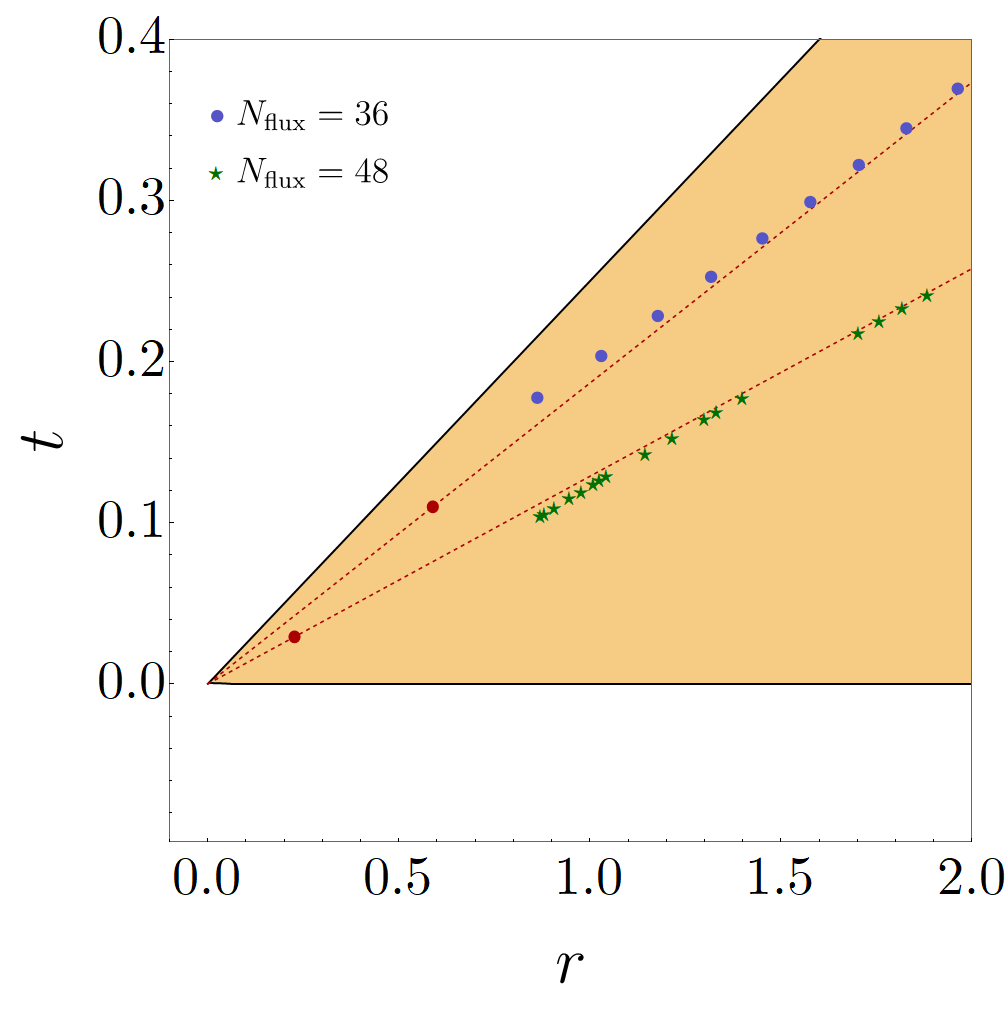}
    \label{fig:cone_true_zoom}
    \end{subfigure}
    \caption{Plots representing two families of solutions obtained from the first-step results of table \ref{tab:two_parameter_LCS}. (a) On the left, representation of the $(r,t)$-plane along with the CS cone shaded in orange and its stretched version with parameter $c=1$ in a darker tone. Full solutions obtained from different values of $t^0_{\rm target}$ are depicted in blue for the $\Nf=36$ case and in green for the $\Nf=48$ one. As expected, the solutions quickly converge towards the red lines. The two black circles pinpoint the specific solutions worked out in  subsection \ref{sec:full_sols}.  (b) On the right, zoom on the area delimited by the dashed rectangle on the left figure. Solutions close to the tip of the CS cone (and far from the stretched cone) feature small  deviations from the first-step values $v^i_{(0)}$.}
    \label{fig:cone_full}
\end{figure}

This picture is precisely what is observed after applying the complete strategy of subsection \ref{sec:algorithm} to generate solutions, with $t^0_{\rm target}$ chosen in the range $(1, t^0_{\rm max}) \sim (1,800)$ for the $\Nf = 36$ family of solutions, and $(1, t^0_{\rm max}) \sim (1,231)$ for the $\Nf = 48$ family.  For the two-parameter reduction at hand, the true CS cone can be faithfully represented in two dimensions in the $(r,t)$-plane. Figure \ref{fig:cone_full} shows the locations of the solutions and how they fit inside the CS cone and its stretched version. The solutions reach the stretched cone at $t^0_{\rm min}\sim 9.1$ and $t^0_{\rm min}\sim 34.2$ for the $\Nf=36$ and $\Nf=48$ families respectively.  Note that outside the stretched cone, some solutions might be viable as well, since exponential corrections could be small even with $c<1$. The discretum of values along each ray represents different choices for those fluxes not involved in $\Nf$, that are tuned to reach a value as close as possible to $Q'_{\rm target}$ as defined in \eqref{eq:Qptarget}. In fact, a lot of different flux choices can lead to identical values for $Q'$ in \eqref{eq:Qp_singular}. Therefore, in figure \ref{fig:cone_full}, the points depicted in the $(r,t)$-plane are  highly degenerate, in the sense that they describe several solutions that can be obtained from a large number of different choices for the fluxes not involved in $\Nf$.

\subsubsection{Scalar mass spectra}
\label{sec:mass_spectra}

For the two specific solutions described in sect.~\ref{sec:full_sols}, as it has already been explained, we know that all axions are stabilized for the $\Nf=36$ solutions while $11$ directions remain flat in the $\Nf=48$ one. On the other hand, the saxions are expected to be fully stabilized since they are all involved in the non-linear saxionic system of equations. One can however check these claims explicitly by computing numerically the scalar mass spectrum for the vacua under consideration. The easiest way to get the masses is to follow the strategy outlined in \cite{Blanco-Pillado:2020wjn}. We compute the matrix $Z$ defined as $Z_{AB}\equiv e^{K/2}D_AD_BW$ and evaluate it numerically for the solution at hand. Besides, we evaluate the Kähler metric and compute its inverse, in order to obtain canonically normalized masses in the end. With these two quantities, we can evaluate $Z_{AB}K^{B\bar C}\bar Z_{\bar C\bar D}K^{D\bar C}$, whose eigenvalues give the fermion masses $m_\lambda^2$. Then, we compute the gravitino mass $m_{3/2}\equiv e^{K/2}|W|$ and express the scalar masses squared as
\begin{equation}
\mu_{\pm\lambda}^2=(m_{3/2}\pm m_\lambda)^2\ ,\qquad \lambda=0,\dots,h^{2,1}\ .
\end{equation}
In the $\Nf=36$ vacuum, we find that all masses are $\mathcal{O}(1-100)$ in Planck units except one mode which is very light (see appendix \ref{sec:sol_36} for the detailed spectrum). This light mode is fully expected in the IIB1 scenario from general grounds when a vacuum is close to the LCS point \cite{Marchesano:2021gyv,Coudarchet:2022fcl}. It is related to the fact that we can absorb $t^0$ in the saxionic system of vacuum equations at the LCS point, such that there is a massless mode there. Then, it is natural to expect vacua quite close to this point to feature a light mode. For the $\Nf=48$ vacuum, the numerical computation confirms the presence of $11$ massless modes while all others are $\mathcal{O}(1-100)$ in Planck units except one that is light, as expected.

When computing the mass spectra, we are lead to implement numerically the full prepotential with the $52$ complex variables as well as the full superpotential. We have thus checked explicitly, without any massaging of the vacuum equations, that the solutions we propose (flux configurations along with axionic and saxionic vevs) are indeed flux vacua.

\section{A lower tadpole in a four-parameter reduction}
\label{sec:four-param}

The minimum magnitude of the flux-induced tadpole for solutions arising from the two-parameter reduction of the model has been shown to be $\Nf=36$. This is compatible with the generic estimate \eqref{eq:boundNf}, but it overshoots the tadpole of the explicit orientifold quotient that corresponds to \eqref{TD3basic}. Motivated to find new solutions with tadpoles compatible with this bound, in this section we explore a more complex reduction to four parameters. Notice that playing with fewer flux quanta in the two-parameter case is what enabled us to efficiently explore small-tadpole configurations, so that increasing the number of parameters seems to go in the wrong direction.  However a four-parameter reduction, which is  still a great improvement over $51$, allows to explore vacua with a more involved flux structure.  There is thus hope that we can uncover smaller tadpoles in this setup by benefiting from the structure of the vacuum equations, and without introducing too many variables. We will see that this is indeed the case, and present a solution with $\Nf=32$ found via a four-parameter ansatz. Given the similarities with the two-parameter solutions we will be more sketchy in our discussion, leaving some  numerical details for appendix \ref{sec:sol_32}.

\subsection{The ansatz}

The model defined in section \ref{sec:orb_51} admits three different prescriptions which reduce it to four effective complex structure moduli. One such possible choice for a four-parameter description of the model is given by
\begin{align}
\begin{aligned}
t^i\equiv t^0v^i &\equiv (r,r,r',\underbrace{t,\dots,t}_{\text{32 times}},\underbrace{t',\dots,t'}_{\text{16 times}})\ , \\[5pt]
v^i &\equiv (v^r,v^r,v^{r'},\underbrace{v^t,\dots,v^t}_{\text{32 times}},\underbrace{v^{t'},\dots,v^{t'}}_{\text{16 times}})\ , \\[5pt]
f_A^i &\equiv (f_A^r,f_A^r,f_A^{r'},\underbrace{f_A^t,\dots,f_A^t}_{\text{32 times}},\underbrace{f_A^{t'},\dots,f_A^{t'}}_{\text{16 times}})\ , \\[5pt]
h_i^B &\equiv (h_r^B, h_r^B, h_{r'}^B, \underbrace{h_t^B,\dots,h_t^B}_{\text{32 times}},\underbrace{h_{t'}^B,\dots,h_{t'}^B}_{\text{16 times}})\ . 
\end{aligned}
\label{eq:4_par_ansatz}
\end{align}
Two other ans\"atze that lead to a four-parameter description can be constructed from
\begin{align}
t^i\equiv t^0v^i &\equiv (r,r',r,\underbrace{t,\dots,t}_{\text{16 times}},\underbrace{t',\dots,t'}_{\text{16 times}},\underbrace{t,\dots,t}_{\text{16 times}})\ ,
\end{align}
and
\begin{align}
t^i\equiv t^0v^i &\equiv (r',r,r,\underbrace{t',\dots,t'}_{\text{16 times}},\underbrace{t,\dots,t}_{\text{32 times}})\ ,
\end{align}
in such a way that the vectors $v^i$, $f_A^i$ and $h_i^B$ have the same structure of repeated entries. All of the equations to solve and volumes to check that follow from these reductions are exactly the same, so a solution described with certain saxions $\lbrace r,r',t,t' \rbrace$ and flux components $\lbrace f_A^r,f_A^{r'},f_A^t,f_A^{t'}, h_r^B, h_{r'}^B,  h_t^B, h_{t'}^B\rbrace$ will represent three different potential vacuum families. 

With such an ansatz, the various volumes of the mirror geometry in \eqref{eq:vol_first}-\eqref{eq:vol_last} are more complex than in the two-parameter setup. In particular, it is not possible to summarize the stretched CS cone conditions into a set of few simple inequalities. This is no big deal and it just means that when checking the candidate solutions at the LCS point in the first step of our algorithm, we need to go through all volumes and check if they are positive. With these starting points, the idea will then be the same as before, illustrated in fig.~\ref{fig:cone}, to tune the remaining free fluxes to generate full solutions inside the stretched cone.

\subsection{A new solution}

In the same way as in the previous section for the two-parameter case, we follow here the four steps of our strategy to find solutions and display the results.

\subsubsection{LCS system}

\paragraph{Step 1:} We generate tuples $\{f_A^r,f_A^{r'},f_A^t,f_A^{t'},h_r^B,h_{r'}^B,h_t^B,h_{t'}^B\}$ and restrict to configuration yielding a flux-induced contribution to the tadpole lower than $32$. For each tuple, we solve the LCS system \eqref{eq:saxion_syst_LCS} to find 
$v^r_{(0)}$, $v^{r'}_{(0)}$, $v^t_{(0)}$ and $v^{t'}_{(0)}$, and we plug the solutions into the volumes to test if they are positive. We found only one configuration passing this filter,\footnote{As a crosscheck, we also looked for solutions with larger  tadpoles and, among others, we recovered as a special case the two solutions presented in the previous section in the two-parameter case.} described in table 4.

\setlength\tabcolsep{3.8pt}
\begin{table}
\centering
\begin{tabular}{c|cccccccc|cccc|c|c}
$N_{\text{flux}}$ & $f_A^r$& $f_A^{r'}$ & $f_A^t$ & $f_A^{t'}$ & $h_r^B$ & $h_{r'}^B$ & $h_t^B$ & $h_{t'}^B$ & $v^r_{(0)}$ & $v^{r'}_{(0)}$ & $v^t_{(0)}$  & $v^{t'}_{(0)}$ & $\rank (M)$ & $(t^0)^3\,|\xi|$ \\ \hline
32 & -6 & 4 & 0 & -2 & 2 & -2 & 0 & 0 & 0.438 & 0.391 & 0.086 & 0.061 & 43 & 0.286
\end{tabular}
\caption{A solution with positive volumes found for the four-parameter model at LCS.}
\label{tab:four_parameter_LCS}
\end{table}

\subsubsection{Remaining fluxes and solution}

We proceed with the rest of the strategy to generate full solutions. The rank of the matrix $M$ for the case at hand is not maximal such that some axions are left unstabilized by \eqref{eq:MB} and we will have to take care about flux constraints in the following steps.

\paragraph{Step 2:} Note that again, the small value for $(t^0)^3\,|\xi|$ ensures that $t^0$ needs not be very large to achieve $|\xi|\ll 1$, such that the strongest constraints come from the volumes. More precisely, we take $t^0_{\rm target}=22$ and from the values displayed in table.~\ref{tab:four_parameter_LCS} we obtain $Q'_{\rm target}=-0.113$ using \eqref{eq:Qptarget}.

\paragraph{Step 3:} We now want to fix all the remaining flux quanta $f_0^B$ $f_i^B$ and $h_0^B$ by minimizing $(Q'-Q'_{\rm target})^2$ over the integers, with $Q'$ expressed in terms of fluxes with \eqref{eq:Qp_singular}, which is valid when $M$ is not invertible. The $53$ fluxes are however not independent due to the constraints arising from the axionic system. These constraints can be shown to be
\begin{equation}
\begin{aligned}
&f^B_{36}-f^B_{39}-f^B_{48}+f^B_{51}=0\ ,\qquad\quad &&f^B_{36}-f^B_{38}-f^B_{48}+f^B_{50}=0\ ,\\
&f^B_{36}-f^B_{37}-f^B_{48}+f^B_{49}=0\ , &&f^B_{36}-f^B_{39}-f^B_{44}+f^B_{47}=0\ ,\\
&f^B_{36}-f^B_{38}-f^B_{44}+f^B_{46}=0\ , &&f^B_{36}-f^B_{37}-f^B_{44}+f^B_{45}=0\ ,\\
&f^B_{36}-f^B_{39}-f^B_{40}+f^B_{43}=0\ , &&f^B_{36}-f^B_{38}-f^B_{40}+f^B_{42}=0\ ,\\
&f^B_{36}-f^B_{37}-f^B_{40}+f^B_{41}=0\ .
\end{aligned}
\end{equation}
A flux set that minimizes $(Q'-Q'_{\rm target})^2$ and satisfies the constraints above can be found in \eqref{eq:fluxesb_32}, which gives a definitive value $Q'=-0.099$.

\paragraph{Step 4:} Plugging all the flux quanta in the full system of equations, we find the definitive solution displayed in table \ref{tab:full_sols_4}.

\setlength\tabcolsep{5pt}
\begin{table}[ht!]
\centering
\begin{tabular}{c|cccccccc|c|ccccc}
$N_{\text{flux}}$ & $f_A^r$ & $f_A^{r'}$ & $f_A^t$ & $f_A^{t'}$ & $h_r^B$ & $h_{r'}^B$ & $h_t^B$  & $h_{t'}^B$ & $f_0^B,f_i^B,h_0^B$ & $t^0$ & $r$ & $r'$ & $t$ & $t'$\\ \hline
32 & -6 & 4 & 0 & -2 & 2 & -2 & 0 & 0 & See \eqref{eq:fluxesb_32} & 25.1 & 11.0 & 9.80 & 2.16 & 1.53
\end{tabular}
\caption{A full solution with $\Nf=32$.}
\label{tab:full_sols_4}
\end{table}
For this vacuum, the total volume of the mirror Calabi--Yau $\V_{Y_3}$, the volume of the divisors generically denoted $\V_{E,D}$ and the areas generically written $\A$ of the curves generating the Mori cone, that can be read from \eqref{eq:vol_first}-\eqref{eq:vol_last}, are all bigger than $1$ and read (with duplicates removed):
\begin{equation}
\V_{Y_3}=1076\ ,\ \ \V_{E,D}=(24.1,8.24,5.70,5.53)\ ,\ \  \A=(4.88,2.35,1.17,1.53,2.79)\ .
\end{equation}

\subsubsection{Mass spectrum}

Applying the same techniques as described in sect.~\ref{sec:mass_spectra}, we can evaluate the scalar masses at the vacuum. As expected, we find $9$ massless modes corresponding to the unfixed axions due to the rank of $M$ being only $43$. One mode is very light, as expected for solutions close to the LCS point in the IIB1 setup \cite{Coudarchet:2022fcl}. Eventually, the other masses are of order $\mathcal{O}(0.01-100)$ in Planck units. The full list of masses is displayed in \eqref{eq:masses_num}.

\subsection{Other, more complicated reductions}

So far we have explored reductions to effectively two and four complex structure moduli. However, a six-parameter reduction is also possible, whose main properties are presented in appendix \ref{sec:six-param}. In that particular case, the higher number of variables and flux quanta makes the numerical search more complicated (though still tractable). Regrettably, the numerical search over $10^7$ potential flux tuples with $\Nf\leq 32$ has not led to any new solution which had not already been found in the aforementioned reductions. 

\section{Connection with the Tadpole Conjecture}

The solutions described above stabilize a moderately large number of real moduli ($2h^{2,1}+2=104$ when $M$ has maximal rank) at LCS, while having a relatively small flux-induced D3-brane charge. It is thus interesting to see how this relates to the Tadpole Conjecture \cite{Bena:2020xrh}, whose features we review in what follows. 

\subsection{Versions of the conjecture}

Initially proposed in \cite{Bena:2020xrh}, the Tadpole Conjecture applies to M-theory, F-theory or type IIB 
Calabi--Yau compactifications with fluxes. Its general form is the following:
\begin{equation}
\Nf>\alpha\, n_{\rm stab}\quad\text{ for }\quad n_{\rm stab}\gg 1\quad\text{ with }\quad \alpha=\mathcal{O}(1)\ ,
\label{TC}
\end{equation}
where $n_{\rm stab}$ counts the numbers of moduli that are stabilized by fluxes.  The refined version of the conjecture \cite{Bena:2020xrh} further predicts a lower bound on the so-called {\it flux-tadpole constant} $\alpha$, namely $\alpha>1/3$. In the F-theory context, such a linear growth with coefficient $1/3$ of the flux-induced tadpole in the number of moduli leads to a no-go for stabilizing a large number of moduli without overshooting the tadpole upper bound set by the Euler characteristic of the fourfold, estimated to also grow linearly but with a smaller coefficient $1/4$. A more recent refinement distinguishes between a weak and a strong form of the conjecture \cite{Lust:2022mhk}. The strong form states that the linear scaling of $\Nf$ with the number of stabilized moduli $n_{\rm stab}$ holds, no matter what this number is. On the contrary, the weak form proposes that the linear scaling is to be observed only when achieving full moduli stabilization in a given compactification.

In the following we will use our results to test the strong version of the Tadpole Conjecture, with particular interest on the ratio $\Nf/n_{\rm stab}$, that measures how efficient the fluxes are in stabilizing moduli. Because we mostly care about non-supersymmetric vacua, it makes more sense that $n_{\rm stab}$ counts the number of {\em real} moduli stabilized by fluxes. Then, in order to match the conventions for $\a$ in  \cite{Bena:2020xrh}, one needs to compute the flux tadpole $\Nf$ in the covering space, as we have done in \eqref{eq:Nflux}.

\subsection{Regimes of validity}

In the quest to prove or challenge the Tadpole Conjecture, two kinds of analysis have recently tested the proposal in two very different regimes, obtaining results that naively seem to point in opposite directions. On the one hand, the proposal has been tested in a large class of asymptotic regimes in complex structure moduli space  \cite{Grimm:2021ckh,Grana:2022dfw}, finding a strong support of \eqref{TC} that could a priori hint towards a general, mathematical proof of this statement. In fact, in such asymptotic regimes, one can show that the flux-induced contribution to the D3-brane tadpole is bounded from below by
\begin{equation}
\text{Strict asymptotic regime: }\quad\Nf>0.7\, n_{\rm stab}\ ,
\end{equation}
which is a stronger statement than the conjectured $\a=1/3$ value. It is important to notice that this derivation applies to a specific domain of validity, as it assumes a \emph{strict asymptotic regime}, where not only the complex structure moduli grow towards corners in moduli space, but they do so by following a strict hierarchy among them. 

On the other hand, far away from this regime and deep inside the interior of complex structure moduli space, vacua in F-theory at Fermat points were found to violate the Tadpole Conjecture in its strong form \cite{Lust:2022mhk}. There, an example has been worked out with a ratio of the flux-induced tadpole to the number of stabilized moduli satisfying
\begin{equation}
\text{A construction at a Fermat point: }\quad\frac{\Nf}{n_{\rm stab}}=0.003\ ,    
\end{equation}
which is in strong tension with the $\a =1/3$ value and even the $\mathcal{O}(1)$ estimate in \eqref{TC}.

\begin{figure}[p]
    \centering
    \includegraphics[scale=0.62]{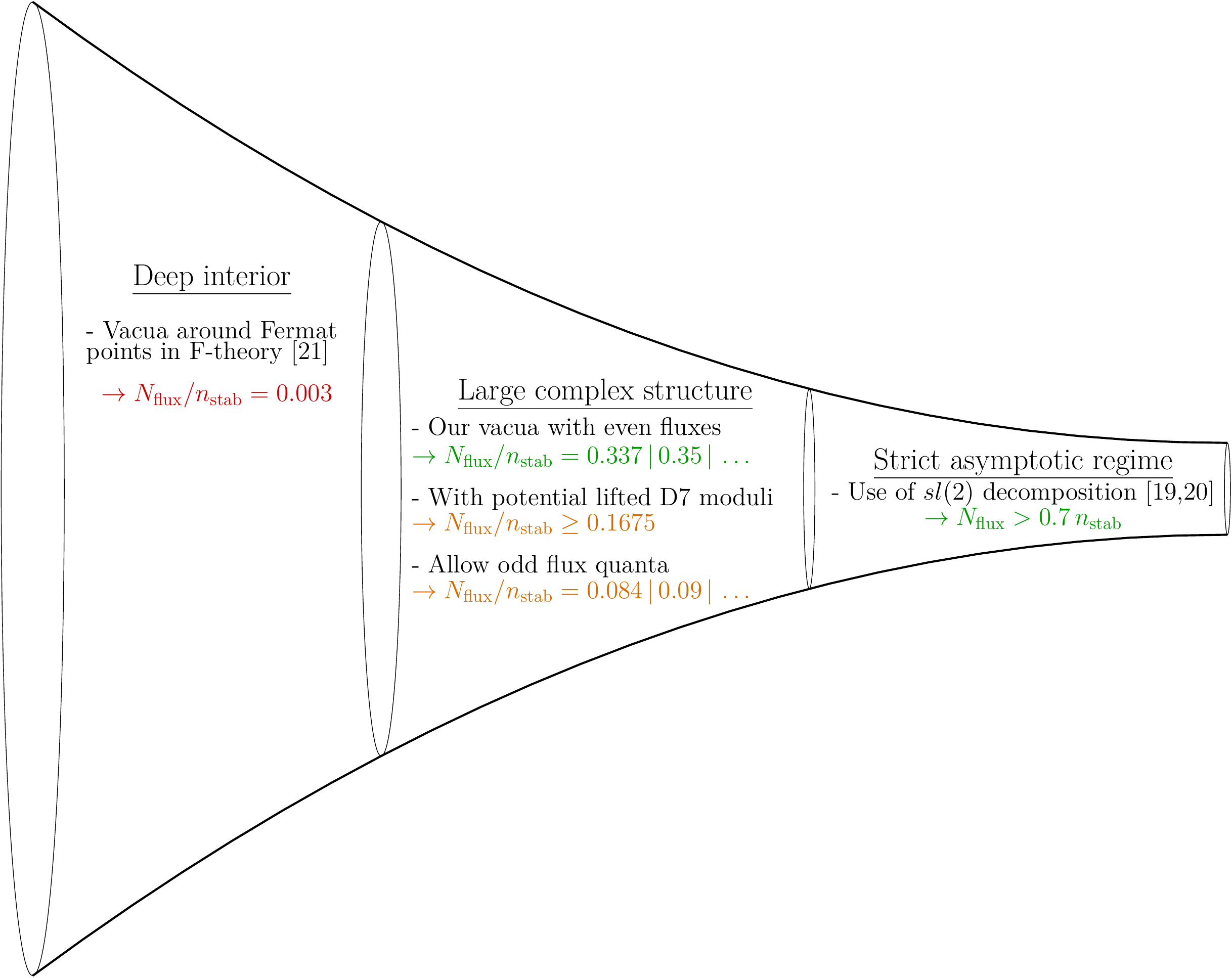}
    \caption{An explicit F-theory construction around Fermat points, deep inside the interior of moduli space, displays a very small $\Nf /n_{\rm stab}$ ratio \cite{Lust:2022mhk}, in tension with the conjectured $1/3$ bound. On the other side of moduli space, in the strict asymptotic regime, a lower bound $0.7$ for this ratio has been obtained in \cite{Grimm:2021ckh,Grana:2022dfw}. In between, in large complex structure regimes with symmetric fluxes, we have found solutions with $\mathcal{O}(100)$ stabilized moduli and moderate flux-induced tadpoles.   Assuming O3$^-$-planes and imposing even flux quanta, the solutions we obtain are in agreement with the refined Tadpole Conjecture, with a slope very close to the $\a = 1/3$ bound  in the case of our $\Nf=32$ solution. As a consequence, any additional modulus stabilized in the D7-brane sector would violate the bound. An upper estimate on the number of these D7-brane moduli gives  $\a \sim 1/6$, in clear tension with the bound of the refined conjecture.  More speculative solutions, requiring the presence of exotic O$3^+$-planes compatible with odd flux quanta show a much smaller ratio $\alpha\sim 1/12$, again in tension with the conjectured bound.}
    \label{fig:conjecture}
\end{figure}

If one concludes that the strongest form of the Tadpole Conjecture is proven true in the strict asymptotic regime and proven wrong in the deep interior of moduli space, a natural question is to determine where lies the switch between validity and non-validity of such a proposal. In this paper, we have investigated the LCS regime, which is a particular case of asymptotic regime but less stringent than a strict limit, because it does not need to lie in any of the growth sectors considered in \cite{Grimm:2021ckh,Grana:2022dfw}. In fact, in our effective reductions there is a large symmetry among flux quanta that results in a lot of vevs being identical, somewhat analogously to the symmetric points studied in \cite{Lust:2022mhk}. Therefore, they are asymptotic regimes that lie far away from the hierarchy of saxionic vevs present in the growth sectors analyzed in \cite{Grimm:2021ckh,Grana:2022dfw}. From this point of view, our analysis lies in between the two regimes explored in \cite{Grimm:2021ckh,Grana:2022dfw} and \cite{Lust:2022mhk}, and one may hope   to measure how this intermediate regime interpolates between the other two in terms of the value for $\alpha$, as illustrated by fig.~\ref{fig:conjecture}.

The most solid of our solutions are those obtained in section \ref{sec:four-param}, with $\Nf=32$ and compatible with the explicit orientifold projection leading to \eqref{TD3basic}, followed by those of section \ref{sec:two-param}, with $\Nf=36$ and $48$ and compatible with the estimate \eqref{eq:boundNf}. These examples show ratios of the flux-induced tadpole to the number of stabilized moduli as small as
\begin{equation}
\text{Solutions at LCS with even fluxes: }\quad\frac{\Nf}{n_{\rm stab}}= \begin{cases}0.3368 \\ 0.35 \\ 0.52\end{cases},
\end{equation}
respectively from top to bottom. These values are compatible with the $\a=1/3$ value of the refined conjecture, but are much closer to this value than the $0.7$ bound derived in the strict asymptotic regime. They are displayed in green in the LCS region on fig.~\ref{fig:conjecture}. Even if they are in agreement with the strongest form of the Tadpole Conjecture, note that these solutions are in direct tension with the expectations raised in \cite{Plauschinn:2021hkp,Lust:2021xds,Tsagkaris:2022apo} in the context of type IIB flux compactifications, where it is argued that by going deeper inside the CS cone, the smallest value for $\Nf$ should increase. Indeed, our algorithm allows us to generate solutions inside a very stretched CS cone by varying the  value of $t^0_{\rm target}$ without changing the flux-induced contribution to the tadpole.

\subsection{Adding D7-brane moduli}

Our counting for $n_{\rm stab}$ above only involves closed-string fields, namely the axio-dilaton and complex structure moduli. This is motivated by the type IIB version of the Tadpole Conjecture proposed in \cite{Bena:2020xrh}, which has separate statements for bulk and for D7-bane position moduli, even if both sets are quite similar from an F-theory perspective. The reason behind such a separation seems to be the assumption that each set of moduli is fixed by a different set of fluxes: bulk moduli are fixed by three-form fluxes, while D7-brane positions by their worldvolume fluxes.

While D7-brane worldvolume fluxes do generate a superpotential for their positions, see e.g. \cite{Jockers:2005zy,Martucci:2006ij}, bulk three-form fluxes are known to stabilize open string moduli as well. In particular, they generically stabilize D7-brane position moduli even in the absence of any worldvolume flux \cite{Gorlich:2004qm,Camara:2004jj,Gomis:2005wc}.\footnote{See \cite{Landete:2017amp} for the interplay between closed and open string fluxes in a simple F-theory setup, and \cite{Marchesano:2014iea} for that between three-form fluxes and D7-brane Wilson line moduli stabilization.} In a setup with D7-branes on top of O7-planes without any worldvolume flux, it thus makes sense to include their position moduli in the computation of $n_{\rm stab}$, because one is fixing more moduli at no tadpole cost. It is the computation of the ratio $\Nf /n_{\rm stab}$ with this definition of $n_{\rm stab}$ that should match the one defined from the F-theory perspective. 

Let us for instance consider the choice of flux vacua of section  \ref{sec:four-param} with $\Nf =32$. This model is compatible with the Calabi--Yau deformation of the $T^6/(\Z_2\times\Z_2)$ orientifold T-dual to \cite{Berkooz:1996dw}. As discussed in section \ref{sec:quantization}, in the orbifold limit one may cancel the D7-brane tadpole by placing 4 D7-branes plus their orientifold images on top of each of the 12 O7$^-$-planes, without the need of any worldvolume flux on them. This gives the tadpole \eqref{TD3basic}, and a total of 48 complexified D7-brane position moduli. 

Upon deformation to the $(3,51)$ Calabi--Yau orientifold, one expects the four-cycles wrapped by the O7-planes to change their embedding and possibly recombine among themselves. However, by continuity one should be able to still cancel their tadpole by placing D7-branes on top of them, and in such a way that \eqref{TD3basic} is maintained. In particular, this means that neither the O7-plane nor the D7-brane curvature terms are such that they contribute to the D3-brane tadpole, as we have assumed in our analysis. 
 One then deduces that the Euler characteristic of these divisors vanishes, which typically corresponds to a topology of the form $T^2 \times \Sigma_2$, with $\Sigma_2$ a Riemann surface. Except for the case $\Sigma_2 = \mathbb{P}^1$, this means that the D7-branes wrapped on these divisors will have position moduli and, since we precisely are in the setup of \cite{Gomis:2005wc}, we expect such moduli to be lifted by the three-form background fluxes. 
 
A detailed computation of the lifted D7-brane moduli would require a more detailed analysis of the O7-plane divisors in the $(3,51)$ Calabi--Yau orientifold that is beyond the scope of this work. However, based on the generality of the D7-brane moduli-fixing mechanism of \cite{Gomis:2005wc} and the freedom in the choice of flux quanta that lead to the same tadpole, one can assume that any D7-brane position moduli in the $(3,51)$ Calabi--Yau orientifold should be fixed. Then, given the number of D7-brane position moduli in the orbifold limit, one can estimate that the contribution to $n_{\rm stab}$  could be as large as 96. Combined with our previous discussion, one finds that for the flux vacua of section \ref{sec:four-param} the  ratio $\Nf/n_{\rm stab}$ should be in the range
\begin{equation}
0.1675 \ \leq \ \frac{\Nf}{n_{\rm stab}} \ \leq \  0.3368 \ ,
\label{D7count}
\end{equation}
which essentially challenges the proposed value of $1/3$. Indeed, notice that with just a single D7-brane position stabilized by fluxes one recovers $\a < 1/3$ in \eqref{TC}, and therefore counterexamples to the refined Tadpole Conjecture. It would be interesting to confirm this expectation from a more detailed analysis, as well as from an F-theory perspective. 

The same considerations can be applied to the vacua of section \ref{sec:two-param} and their F-theory lift. Of course, in this case the result depends on the choice of orientifold projection leading to a D3-brane tadpole within the estimate \eqref{eq:boundNf}, although it is easy to see that as long as there are D7-branes with position moduli, the same tension with the bound $\a = 1/3$ will exist, at least for the set of flux vacua with $\Nf = 36$. 

An interesting caveat of the above discussion is whether the D7-brane sector gives rise to non-Abelian gauge groups, that in turn lead to singularities in the F-theory uplift of these vacua. Indeed, in the orbifold limit of the $(3,51)$ Calabi--Yau orientifold, pairs of D7-branes host $SU(2)$ gauge groups, independently of their position. It is not clear whether this feature remains after a deformation away from the orbifold limit, although of course one expects that several D7-branes placed on top of an O7-plane host a non-Abelian gauge group. In this sense, one should stress that the moduli fixing mechanism of \cite{Gomis:2005wc} also leads to a discretum of D7-brane positions away from the O7-plane locus, so such a non-Abelian enhancement need not take place. Again, a detailed study of the 7-brane sector would be needed in order to have a definite picture. 

\subsection{Odd flux quanta}

As mentioned in sect.~\ref{sec:quantization}, we imposed all flux quanta to be even to be consistent with flux quantization \cite{Frey:2002hf} in the presence of only O$3^-$-planes. In order to have odd flux quanta, one must instead consider compactifications with both O$3^-$-planes and O$3^+$-planes. These latter, more exotic O-planes are the consequence of the presence of a discrete background for the antisymmetric B-field \cite{Bianchi:1991eu,Bianchi:1997rf,Witten:1997bs,Blumenhagen:2000ea}. More precisely, a flux quantum must be odd if the associated 3-cycle crosses an odd number of exotic planes \cite{Frey:2002hf,Cascales:2003zp}. However, since the O$3^+$-planes have positive tension and charge, exchanging regular planes for exotic ones will increase the D3-charge $Q_{\rm D3}$ to less negative values and thus lower the upper bound on $\Nf$. The consistency of the solutions presented below is thus subject to the existence of an exotic orientifold projection such that the upper bound on the flux-induced tadpole does not drop below the associated values for $\Nf$. The lack of exhaustive information about the orientifold projections we are allowed to consider in our geometry prevents us from precisely investigate if this is indeed possible or not. Nevertheless, we find important to point out that, when we apply our algorithm without forbidding odd fluxes, we find solutions with much lower flux-induced tadpole contributions. Some of these solutions for the step $1$ of the algorithm  and in the two-parameter setup are displayed in table \ref{tab:two_parameter_LCS_odd}. Full solutions, inside the stretched CS cone, can easily be obtained by following the same procedure as before. 

\begin{table}[ht!]
\centering
\begin{tabular}{c|cccc|cc|c|c}
$N_{\text{flux}}$ & $f_A^r$ & $f_A^t$ & $h_r^B$ & $h_t^B$ & $v^r_{(0)}$ & $v^t_{(0)}$ & $\rank (M)$ & $N_{\text{flux}}/n_{\text{stab}}$ \\ \hline
9 & -3 & -1 & 1 & 0 & 0.589 & 0.110 & 52 & 0.087\\
12 & -4 & -1 & 1 & 0 & 0.227 & 0.029 & 41 & 0.129\\
15 & -5 & -2 & 1 & 0 & 0.796 & 0.180 & 52 & 0.144\\
15 & -5 & -1 & 1 & 0 & 0.143 & 0.014 & 52 & 0.144\\
18 & -3 & -1 & 2 & 0 & 1.178 & 0.220 & 52 & 0.173\\
24 & -4 & -1 & 2 & 0 & 0.454 & 0.058 & 41 & 0.258\\
27 & -3 & -1 & 3 & 0 & 1.776 & 0.330 & 52 & 0.260\\
30 & -5 & -2 & 2 & 0 & 1.592 & 0.360 & 52 & 0.288\\
30 & -5 & -1 & 2 & 0 & 0.287 & 0.028 & 52 & 0.288\\
30 & -2 & -1 & -3 & 1 & 3.982 & 0.842 & 25 & 0.390\\
33 & -1 & -1 & -5 & 1 & 1.486 & 0.299 & 52 & 0.317
\end{tabular}
\caption{Some numerical solutions found for the two-parameter model at LCS without forbidding odd fluxes. Each solution is associated with another one featuring opposite fluxes, that we do not write for brevity.}
\label{tab:two_parameter_LCS_odd}
\end{table}

When considering the specific orientifold projection mentioned in sect.~\ref{sec:quantization} which is compatible with the orbifold limit $T^6/(\Z_2\times\Z_2)$, it has been shown that at the orbifold point, the simplest non-trivial B-field background with rank two does not allow for odd fluxes \cite{Cascales:2003zp}. The reason is that the O3-planes distribution is such that any 3-cycle crosses an even number of exotic planes. Things are expected to change and get more involved if the rank of the discrete background is at least four. The D3-charge being divided by the rank, it would decrease down to $-8$. This could still be compatible with the solution containing odd fluxes obtained from the one with $\Nf=32$, presented in table \ref{tab:full_sols_4}, by dividing all quanta by $2$, since the tadpole would also drop to $\Nf=8$.

Subject to all the precautions already mentioned, this $\Nf=8$ solution together with those presented in table \ref{tab:two_parameter_LCS_odd} would yield ratios as small as 
\begin{equation}
\text{Solutions at LCS with odd fluxes allowed: }\quad\frac{\Nf}{n_{\rm stab}}=\begin{cases} 0.084\\ 0.087\\ $\dots$\end{cases}\ ,
\end{equation}
and would therefore provide counterexamples to the refined Tadpole Conjecture at large complex structure. It would be very interesting to see if these options could be realized. More generally, it would be interesting to see if the value for the bound  $\a$ could be dependent on the content of exotic O3-planes.

\section{Conclusions and outlook}

In this paper we applied the \emph{IIB1 scenario}, introduced in \cite{Marchesano:2021gyv}, to a specific compactification geometry. The IIB1 choice of fluxes has been extensively explored in \cite{Coudarchet:2022fcl} and has been shown to provide a promising first-step simplifying scheme for moduli stabilization regardless of the number of scalar fields. One of the interesting features of this flux family is a nice split between axions and saxions at the level of the vacuum equations. More specifically, the axions obey a very simple linear system while the saxions satisfy a set of non-linear relations, decoupled from the axionic vevs. Another interesting property is the way the various flux quanta are involved in the equations. The axionic linear relations, together with the saxionic system approximated at large complex structure, only depend on the fluxes that contribute to the flux-induced D3-brane tadpole. One can thus easily generate flux tuples constrained by the tadpole bound and quickly check if the axions are stabilized, reducing the search for vacua to a $h^{2,1}+1$ variables problem instead of twice that number. The saxionic system remains hard to solve generically if no further simplifications are assumed, but one can already notice that its structure allows to efficiently look for vacua in the LCS regime due to the property outlined above. At the LCS point, the system is $h^{2,1}$-dimensional and, as already said, it only involves the flux quanta constrained by the tadpole. If a solution can be found at this level of approximation, the remaining fluxes can be tuned such that the flux-dependent quantity denoted $Q'$ is small in absolute value, which ensures that the full solution is indeed close to LCS and to the first-order approximation. This provides an efficient way to determine promising flux tuples instead of performing, e.g., a random search.

In this paper we applied the procedure described above in order to search for vacua in a model with a moderately large number of complex structure moduli. The specific Calabi-Yau was chosen due to the structure of the triple intersection numbers of its mirror dual, which allows for an effective reduction of the number of saxions down to only two, four or six variables, hence rendering the saxionic system manageable from a numerical point of view. Such reduction ansätze make profit of the topological symmetries of the mirror dual but are different from proper truncations at symmetry points in the moduli space. The mirror compactification geometry in question was the symmetrically resolved orbifold $T^6/(\Z_2\times \Z_2)$ with $h^{1,1}=51$ \cite{Denef:2005mm,Faraggi:2021uli,Reffert:2006du,Lust:2006zh}.

The lack of exhaustive information about the possible ways to orientifold our geometry, and in particular with projections not compatible with the orbifold limit, lead us to use an estimate for the minimum D3-charge possible \cite{Tsagkaris:2022apo,Collinucci:2008pf,Carta:2020ohw,Crino:2022zjk,Gao:2022fdi} and thus for the maximum allowed flux-induced contribution to the D3-brane tadpole. Besides, a more explicit and stringent bound was derived for a specific orientifold projection consistent with the orbifold limit, studied in \cite{Cascales:2003zp}. Fortunately, the analysis of the flux vacua equations is independent of the orientifold action, provided that the same set of complex structure moduli survive the projection. Therefore, with these two bounds in mind and following our algorithm, we found families of solutions of the approximate vacuum equations at the LCS point, that we then promoted to full solutions in the LCS regime by adequately tuning the other flux quanta that are not involved in the tadpole. At this step, we also ensured that mirror volumes and areas were sufficiently big for our vacua to lie inside the stretched complex structure cone, where the LCS approximation is justified due to negligible exponential corrections. At the end of the day, the solutions we found stabilize all $2h^{2,1}+2=104$ scalars or a large portion of them and feature rather small flux-induced tadpoles.  

We eventually ended our analysis by discussing how our solutions relate to the Tadpole Conjecture \cite{Bena:2020xrh}. Assuming flux quanta to be even due to flux quantization in presence of regular O$3^-$-planes \cite{Frey:2002hf}, the only solutions that survive the filtering are those with a ratio of the flux-induced tadpole to the number of stabilized moduli that is in agreement with the conjectured flux tadpole constant $\alpha=1/3$. Some solutions reach values very close to this bound and are thus smaller than the result $\alpha>0.7$ derived in the strict asymptotic corner of moduli space studied in \cite{Grimm:2021ckh,Grana:2022dfw}. The sharp proximity we found in our solutions of the flux tadpole constant with respect to the refined bound led us to discuss effects that may imply crossing this threshold. Firstly, in addition to the bulk moduli, three-form fluxes also have the potential to fix some of the moduli associated to the position of the D7-branes without requiring the presence of worldvolume fluxes. Accounting for this phenomenon in our solutions, the flux-tapdole-to-stabilized moduli ratio could be decreased as low as $\alpha \sim 1/6$. Determining whether this is actually the case would demand a detailed analysis of the divisors in our compactification geometry, which is beyond the scope of this work. Secondly, if the assumption of even fluxes is relaxed, which would require the existence of consistent orientifoldings compatible with the presence of exotic O$3^+$-planes, solutions with a ratio as low as $\alpha\sim 1/12$ can be found, which are in sharp tension with the conjecture. These solutions are however more speculative since we lack explicit information about the orientifold projections allowed in our geometry and their precise O-plane content.

Our results suggest a picture that can be summarized by fig. \ref{fig:conjecture}. Regions in complex structure moduli space that correspond to strict asymptotic regimes, hence a strict hierarchy of saxions vevs, display a flux-tadpole-to-stabilized-moduli ratio above $0.7$. Removing the hierarchy of vevs but staying in an asymptotic regime like the LCS region, one gets closer to the proposed value for the lower bound for this ratio of $1/3$ in the refined TC, and entering the interior of the moduli space further lowers this bound. It would be very interesting to confirm this picture by analyzing further setups, and see if  a universal behaviour for this ratio can be obtained. If that was the case, one would have a global pattern across CS field space indicating how efficient is this moduli stabilization mechanism by background fluxes. It is also important to stress that the content of  fig. \ref{fig:conjecture} only applies to the strong version of the Tadpole Conjecture, and that for the weak version, where full moduli stabilization is required, a different result could apply. Notice nevertheless that in this work we have found potential solutions which achieve stabilization of \emph{all} the complex structure moduli, for which $\rank(M)$ is maximal, i.e. $52$ (see tables \ref{tab:two_parameter_LCS} and \ref{tab:two_parameter_LCS_odd}). Some of those solutions, when written in terms of even fluxes, yield a tadpole of $\Nf=36$ and feature $\alpha\sim 0.35$. On the other hand, allowing the flux quanta to be odd integers reduces those quantities down to $\Nf=9$ and $\alpha\sim 1/12$. It thus seems that  further investigation along these lines could help to clarify what the final picture is for both versions of the Tadpole Conjecture.

In particular, as already mentioned above, getting more information about the compactification geometry studied in this paper would be very enlightening. In particular, a classification of the possible orientifold projections performed away from the orbifold point (of which not all of them may possess a well-defined orbifold limit), their O$3$-planes content and the associated D3-charge would strengthen the robustness of some solutions uncovered in this paper. Information about the fourfold uplifts and their Euler characteristics could also be valuable to address these questions. Along these lines, it would also be very interesting to directly try to apply the strategy followed in this paper with the equivalent of the IIB1 scenario in F-theory \cite{Marchesano:2021gyv}. Working with a known fourfold in F-theory would settle all the issues regarding the topology of the compactification space raised above. In addition, solutions found this way would clearly stabilize D7-brane moduli in addition to the complex structure fields from the IIB perspective and might decrease the values of the ratio of the flux-induced tadpole to the number of stabilized scalars as mentioned above. 

Eventually, it would be insightful to apply our results and procedures to other compactification spaces, extracted for example from the Kreuzer-Skarke database \cite{Kreuzer:2000xy}, that would allow for effective reductions inside the IIB1 scenario. Additionally, it would be interesting to see if our approach could be extended to other moduli stabilization schemes, like the IIB2 scenario of \cite{Marchesano:2021gyv} and its F-theory version. Indeed, it could be that using symmetric fluxes one could simplify the saxions' equations in order to solve them explicitly, and clarify whether this scenario yields counterexamples to the Tadpole Conjecture. Finally, it would be interesting to extend the computation of the flux-tadpole-to-stabilized-moduli ratio to other flux compactification schemes that lead to non-supersymmetric Minkowski vacua at tree level, like those in \cite{Lust:2008zd,Held:2010az}.

%%%%%%%%%%%%%%%%%%%%%%%%%%%%%%%%%%%%%%%%%%%%%%%%%%%%

\section*{Acknowledgments}

We thank Emilian Dudas, I\~naki Garc\'ia-Etxebarria, \'Alvaro Herráez and Max Wiesner for discussions. This work is supported through the grants EUREXCEL$\_$03 funded by CSIC, CEX2020-001007-S and PID2021-123017NB-I00, funded by MCIN/AEI/10.13039/\\ 501100011033 and by ERDF A way of making Europe, by the Spanish Ministry MCIU/\\AEI/FEDER 
grant (PID2021-123703NB-C21) and by the Basque Government grant (IT-1628-22). D. P. is supported through the grant FPU19/04298 funded by MCIN/AEI/\\10.1\-3039/501100011033 and by ESF Investing in your future. 

%%%%%%%%%%%%%%%%%%%%%%%%%%%%%%%%%%%%%%%%%%%%%%%%%%%%

\appendix
\makeatletter
\DeclareRobustCommand{\@seccntformat}[1]{%
  \def\temp@@a{#1}%
  \def\temp@@b{section}%
  \ifx\temp@@a\temp@@b
  \appendixname\ \thesection:\quad%
  \else
  \csname the#1\endcsname\quad%
  \fi
} 
\makeatother
\renewcommand{\theequation}{A.\arabic{equation}}

\section{Two solutions in detail}
\label{sec:Solutions}

In this appendix, we give all numerical details about two of the full solutions presented in the core of the paper. Namely, the solution arising in the four-parameter reduction with $\Nf=32$ and the one achieving full complex structure stabilization in the two-parameter reduction with $\Nf=36$.

\subsection[The tadpole $32$ solution]{\bm The $\Nf = 32$ solution}
\label{sec:sol_32}

This solutions was found in the four-parameter reduction of our model.

\paragraph{\bm Fluxes $f_A^i$ and $h_i^B$}

\begin{align}
\begin{split}
(f_A^1,\dots,f_A^{51})=&(-6,-6,4,0,0,0,0,0,0,0,0,0,0,0,0,0,0,0,0,0,0,0,0,0,0,0,0,\\
&0,0,0,0,0,0,0,0,-2,-2,-2,-2,-2,-2,-2,-2,-2,-2,-2,\\
&-2,-2,-2,-2,-2)\ ,\\[5pt]
(h_1^B,\dots,h_{51}^B)=&(2,2,-2,0,0,0,0,0,0,0,0,0,0,0,0,0,0,0,0,0,0,0,0,0,0,0,\\
&0,0,0,0,0,0,0,0,0,0,0,0,0,0,0,0,0,0,0,0,0,0,0,0,0)\ .
\end{split}
\end{align}

\paragraph{Constraints and remaining fluxes\\\\}

With these $f_A^i$ fluxes, the matrix $M$ defined in \eqref{eq:MLQ} has a rank of $43$ such that only $43$ axions are stabilized by \eqref{eq:MB} while $9$ constraints arise, as summarized in \eqref{eq:flux_constraints}. The constraints are given by
\begin{equation}
\begin{aligned}
&f^B_{36}-f^B_{39}-f^B_{48}+f^B_{51}=0\ ,\qquad\quad &&f^B_{36}-f^B_{38}-f^B_{48}+f^B_{50}=0\ ,\\
&f^B_{36}-f^B_{37}-f^B_{48}+f^B_{49}=0\ , &&f^B_{36}-f^B_{39}-f^B_{44}+f^B_{47}=0\ ,\\
&f^B_{36}-f^B_{38}-f^B_{44}+f^B_{46}=0\ , &&f^B_{36}-f^B_{37}-f^B_{44}+f^B_{45}=0\ ,\\
&f^B_{36}-f^B_{39}-f^B_{40}+f^B_{43}=0\ , &&f^B_{36}-f^B_{38}-f^B_{40}+f^B_{42}=0\ ,\\
&f^B_{36}-f^B_{37}-f^B_{40}+f^B_{41}=0\ .
\end{aligned}
\end{equation}
The solution presented in the main text then has the following flux choice, which satisfies the constraints above:
\begin{align}
\begin{split}
\label{eq:fluxesb_32}
(f_1^B,\dots,f_{51}^B)=&(-2,-2,-4,2,0,-4,-2,-2,-6,-2,0,-8,2,0,4,0,8,-4,-2,\\
&0,6,-4,-2,0,8,-4,-6,0,-8,2,2,4,6,10,-2,-6,2,2,2,6,14,\\
&14,14,2,10,10,10,2,10,10,10)\ ,\\[5pt]
&\hspace{-2.6cm}f_0^B=2\ ,\qquad h_0^B=6\ .
\end{split}
\end{align}

\paragraph{Axions, saxions and volumes\\\\}

Using \eqref{eq:flux_constraints} and going back to the original basis of axions, we can write all $b^a$ as involving linear combinations of 9 parameters. For the sake of brevity, let us write a particular solution obtained from that system by taking all of the free parameters to zero:
\begin{align}
\begin{split}
(b^0,\dots,b^{51})=&(4.50,-2.94,-2.94,-2.88,-2.23,-0.651,-0.734,-1.32,\\
&-0.609,1.47,-0.609,-1.19,1.39,-0.0260,-0.609,-1.69,\\
&-0.359,-1.28,0.641,0.0573,-0.766,-0.266,1.23,\\
&-0.0156,-1.43,-1.43,0.568,0.318,-1.02,2.98,-0.516,\\
&-1.27,-0.599,0.901,-1.10,1.15,-1.48,-1.16,-0.914,\\
&-0.789,-0.477,-0.164,0.0859,0.211,-0.977,-0.664,\\
&-0.414,-0.289,-1.35,-1.04,-0.789,-0.664)\ .
\end{split}
\end{align}
On the other hand the saxions are given by
\begin{align}
\begin{split}
(t^0,\dots,t^{51})=&(25.1,11.0,11.0,9.80,2.16,2.16,2.16,2.16,2.16,2.16,2.16,2.16,\\
&2.16,2.16,2.16,2.16,2.16,2.16,2.16,2.16,2.16,2.16,2.16,2.16,\\
&2.16,2.16,2.16,2.16,2.16,2.16,2.16,2.16,2.16,2.16,2.16,2.16,\\
&1.53,1.53,1.53,1.53,1.53,1.53,1.53,1.53,1.53,1.53,1.53,1.53,\\
&1.53,1.53,1.53,1.53)\ .
\end{split}
\end{align}
With these values, the total volume of the mirror Calabi--Yau $\V_{Y_3}$, the volume of the divisors generically denoted $\V_{E,D}$ and the areas generically written $\A$ of the curves generating the Mori cone, that can be read from \eqref{eq:vol_first}-\eqref{eq:vol_last}, are (with duplicates removed)
\begin{equation}
\V_{Y_3}=1076\ ,\ \ \V_{E,D}=(24.1,8.24,5.70,5.53)\ ,\ \  \A=(4.88,2.35,1.17,1.53,2.79)\ .
\end{equation}
These volumes and areas are all bigger than one, which should ensure the solution to be safe from exponential corrections.

\paragraph{Mass spectrum\\\\}

The canonically normalized scalar masses $m_a$, $a\in\{1,\dots,104\}$, evaluated numerically at the vacuum are displayed in \eqref{eq:masses_num}. As expected, we observe $9$ exactlty massless modes that correspond to the $9$ unstabilized axions. We also note the presence of a light mode, fully expected from the IIB1 analytics close to the LCS point \cite{Coudarchet:2022fcl}: 
\begin{align}
\begin{split}
(m_1,\dots,m_{104})=&(0,0,0,0,0,0,0,0,0,9.07\times10^{-8},0.0168,0.0168,0.0168,\\
&0.0168,0.0168,0.0168,0.365,0.617,0.94,0.94,0.94,2.73,2.73,\\
&2.73,2.73,2.73,2.73,2.73,2.73,2.73,2.73,2.73,2.73,2.73,\\
&2.73,2.73,2.73,2.73,2.73,4.26,4.26,4.26,4.26,4.26,4.26,\\
&4.26,4.26,4.26,4.26,4.26,4.26,4.26,4.26,4.26,4.26,4.26,\\
&4.26,8.76,9.68,12.9,12.9,12.9,12.9,12.9,12.9,13.8,13.8,\\
&13.8,13.8,13.8,13.8,13.8,13.8,13.8,13.8,20.3,22,22,22,\\
&44.6,57,62.8,62.8,62.8,62.8,62.8,62.8,103,127,135,135,\\
&135,135,135,135,169,169,169,192,200,280,280,280,319)\ .
\label{eq:masses_num}
\end{split}
\end{align}

\subsection[The fully stabilized tadpole $36$ solution]{\bm The fully stabilized $\Nf = 36$ solution}
\label{sec:sol_36}

This solution has been uncovered in the two-parameter reduction of the model.

\paragraph{Fluxes\\\\}

The whole fluxes are
\begin{align}
\begin{split}
(f_A^1,\dots,f_A^{51})=&(-6,-6,-6,-2,-2,-2,-2,-2,-2,-2,-2,-2,-2,-2,-2,\\
&-2,-2,-2,-2,-2,-2,-2,-2,-2,-2,-2,-2,-2,-2,-2,\\
&-2,-2,-2,-2,-2,-2,-2,-2,-2,-2,-2,-2,-2,-2,-2,\\
&-2,-2,-2,-2,-2,-2)\ ,\\[5pt]
(h_1^B,\dots,h_{51}^B)=&(2,2,2,0,0,0,0,0,0,0,0,0,0,0,0,0,0,0,0,0,0,0,0,0,0,0,0,0,0,\\
&0,0,0,0,0,0,0,0,0,0,0,0,0,0,0,0,0,0,0,0,0,0)\ ,\\[5pt]
(f_1^B,\dots,f_{51}^B)=&(4,0,-2,2,-2,-6,-2,0,2,-2,0,-2,10,-2,8,0,8,-2,-2,2,\\
&0,2,-4,4,2,2,-2,0,0,-4,0,0,-4,4,2,-2,-2,-2,2,-4,-2,\\
&0,0,-2,0,2,0,-4,0,2,-4)\ ,\\[5pt]
&\hspace{-2.6cm}f_0^B=4\ ,\qquad h_0^B=-2
\end{split}
\end{align}

\paragraph{Axions, saxions and volumes\\\\}

In this case, the matrix $M$ is invertible so that the axionic vevs are easily obtained from \eqref{eq:MB}:
\begin{align}
\begin{split}
(b^0,\dots,b^{51})=&(-21.6,0.31,0.0952,0.595,-1.91,0.841,-1.08,-0.492,-0.409,\\
&0.841,-1.08,0.00794,1.51,0.258,0.341,-0.575,0.00794,-0.242,\\
&-0.659,0.925,-0.212,-0.462,-0.212,0.371,0.538,0.288,1.04,\\
&1.12,-0.546,-1.3,0.454,-1.46,0.621,0.871,-0.379,-0.796,\\
&-0.712,0.288,1.37,-0.379,-0.712,-0.212,0.371,-0.379,\\
&-0.462,0.0377,0.621,0.371,-0.796,-0.796,-0.212,0.538)\ .
\end{split}
\end{align}
The saxions read
\begin{align}
\begin{split}
(t^0,\dots,t^{51})=&(10.5,6.19,6.19,6.19,1.16,1.16,1.16,1.16,1.16,1.16,1.16,1.16,\\
&1.16,1.16,1.16,1.16,1.16,1.16,1.16,1.16,1.16,1.16,1.16,1.16,\\
&1.16,1.16,1.16,1.16,1.16,1.16,1.16,1.16,1.16,1.16,1.16,1.16,\\
&1.16,1.16,1.16,1.16,1.16,1.16,1.16,1.16,1.16,1.16,1.16,1.16,\\
&1.16,1.16,1.16,1.16)\ .
\end{split}
\end{align}
The relevant volumes and areas of the mirror Calabi--Yau defined in \eqref{eq:vol_curves} then read
\begin{equation}
\V_{Y_3}=566\ ,\ \ \V_{E}=11.6\ ,\ \ \V_D=4.53\ ,\ \ \A_r=2.13\ ,\ \A_t=1.57\ ,
\end{equation}
and they are all bigger than $1$.

\paragraph{Mass spectrum\\\\}

The canonically normalized scalar masses $m_a$, $a\in\{1,\dots,104\}$, evaluated numerically at the vacuum are given below and as expected, all moduli get a mass and there is one light mode:
\begin{align}
\begin{split}
(m_1,\dots,m_{104})=&(9.52\times10^{-6},4.07,4.07,5.71,5.71,5.71,5.71,5.71,5.71,5.71,\\
&5.71,5.71,5.71,5.71,5.71,5.71,5.71,5.71,5.71,5.71,5.71,\\
&5.71,5.71,5.71,5.71,5.71,5.71,5.71,5.71,5.71,12.4,12.4,\\
&12.4,12.4,12.4,12.4,12.4,12.4,12.4,12.5,20.1,20.2,20.2,\\
&20.2,20.2,20.2,20.2,20.2,20.2,20.2,31.6,31.6,31.6,31.6,\\
&31.6,31.6,31.6,31.6,31.6,31.6,31.6,31.6,31.6,31.6,31.6,\\
&31.6,31.6,31.6,31.6,31.6,31.6,31.6,31.6,31.6,31.6,31.6,\\
&31.6,35,35,35,35,35,35,35,35,35,35,35,36,36,64.2,64.3,\\
&194,194,194,194,194,194,194,194,194,194,194,257)\ .
\end{split}
\end{align}

\section{Six-parameter reduction}
\label{sec:six-param}

As explained in the main text, the vacuum equations of motion in our setup enjoy another possible effective reduction down to six parameters thanks to the ansatz
\begin{align}
\begin{aligned}
t^i\equiv t^0v^i &\equiv (r,r',r'',\underbrace{t,\dots,t}_{\text{16 times}},\underbrace{t',\dots,t'}_{\text{16 times}},\underbrace{t'',\dots,t''}_{\text{16 times}})\ , \\[10pt]
v^i &\equiv (v^r,v^{r'},v^{r''},\underbrace{v^t,\dots,v^t}_{\text{16 times}},\underbrace{v^{t'},\dots,v^{t'}}_{\text{16 times}},\underbrace{v^{t''},\dots,v^{t''}}_{\text{16 times}})\ , \\[10pt]
f_A^i &\equiv (f_A^r,f_A^{r'},f_A^{r''},\underbrace{f_A^t,\dots,f_A^t}_{\text{16 times}},\underbrace{f_A^{t'},\dots,f_A^{t'}}_{\text{16 times}},\underbrace{f_A^{t''},\dots,f_A^{t''}}_{\text{16 times}})\ , \\[10pt]
h_i^B &\equiv (h_r^B, h_{r'}^B, h_{r''}^B, \underbrace{h_t^B,\dots,h_t^B}_{\text{16 times}},\underbrace{h_{t'}^B,\dots,h_{t'}^B}_{\text{16 times}},\underbrace{h_{t''}^B,\dots,h_{t''}^B}_{\text{16 times}})\ . 
\end{aligned}
\label{eq:6_par_ansatz}
\end{align}
Investigating this more generic setup was numerically more challenging but the use of computer clusters helped us scanning a lot of flux configurations. However, despite the several tens of millions of candidate configurations explored, with $\Nf\leq 32$, none gave a satisfying solution apart from the already known four-parameter solution presented in sect.~\ref{sec:four-param}, which as expected arises as a particular subcase. 

\bibliography{refs}

\end{document}